\documentclass[a4paper,11pt]{article}
\pdfoutput=1 

\usepackage{jheppub} 

\usepackage[T1]{fontenc} 

\pdfoutput=1 
\usepackage{graphicx} %
\usepackage{amsmath,amsfonts,amssymb,slashed,float}
\usepackage{hyperref}
\usepackage[normalem]{ulem}
\usepackage{bbold,wasysym}
\usepackage{array,multirow}
\usepackage[utf8]{inputenc}
\usepackage{xspace}
\usepackage{placeins}
\usepackage{csquotes}
\usepackage{setspace}
\usepackage{tocloft}
\newcommand{\AP}{Alcock-Paczy\'nski}

\usepackage[dvipsnames]{xcolor}

\usepackage[noabbrev,capitalize]{cleveref}

\title{BAO+BBN revisited -- Growing the Hubble tension with a 0.7km/s/Mpc constraint}

\author[a,1]{N. Sch\"oneberg,\note{Corresponding author.}}
\author[a,b]{L. Verde,}
\author[a,c]{H. Gil-Mar\'in,}
\author[a,c,d]{S. Brieden}

\affiliation[a]{Institut de Ci\`encies del Cosmos, Universitat de Barcelona, Mart\'{\i} i Franqu\`es 1, Barcelona E08028, Spain}
\affiliation[b]{ICREA, Pg. Llu\'is Companys 23, Barcelona E08010, Spain}
\affiliation[c]{Institut d’Estudis Espacials de Catalunya (IEEC), Barcelona E08034, Spain}
\affiliation[d]{Dept. de  F\'isica Qu\`antica i Astrof\'isica, Universitat de Barcelona, Mart\'i  i Franqu\`es 1, E-08028 Barcelona, Spain.}
\emailAdd{nils.science@gmail.com}

\abstract{The combination of Baryonic Acoustic Oscillation (BAO) data together with light element abundance measurements from Big Bang Nucleosynthesis (BBN) has been shown to constrain the cosmological expansion history to an unprecedented degree. Using the newest LUNA data and DR16 data from SDSS, the BAO+BBN probe puts tight  constraints  on the Hubble parameter ($H_0 = 67.6 \pm 1.0 \mathrm{km/s/Mpc}$), resulting in a $3.7\sigma$ tension with the local distance ladder determination from SH0ES in a $\Lambda$CDM model. In the updated BAO data the high- and low-redshift subsets are mutually in excellent agreement, and there is no longer a mild internal tension to artificially enhance the constraints. Adding the recently-developed ShapeFit analysis yields $H_0 = 68.3 \pm 0.7 \mathrm{km/s/Mpc}$ ($3.8 \sigma$ tension). For combinations with additional data sets, there is a strong synergy with the sound horizon information of the cosmic microwave background, which leads to one of the tightest constraints to date, $H_0 = 68.30\pm 0.45\mathrm{km/s/Mpc}$, in $4.2\sigma$ tension with SH0ES. The region preferred by this combination is perfectly in agreement with that preferred by ShapeFit. The addition of supernova data also yields a $4.2\sigma$ tension with SH0ES for Pantheon, and a $3.5\sigma$ tension for PantheonPLUS. Finally, we show that there is a degree of model-dependence of the BAO+BBN constraints with respect to early-time solutions of the Hubble tension, and the loss of constraining power in extended models depends on whether the model can be additionally constrained from BBN observations.}

\newcommand{\cons}[3]{\ensuremath{#1^{+#2}_{-#3}}}

\begin{document} 
\maketitle
\flushbottom
\enlargethispage*{2\baselineskip}
\section{Introduction}\label{sec:intro}
With current state-of-the art data, there is a significant tension within the $\Lambda$CDM model between the Hubble parameter, $H_0$\,, as measured by the local distance ladder method from SH0ES \cite{Riess:2021jrx} (but not limited to SH0ES, see also \cite{Freedman:2019jwv,Freedman:2020dne,Anand:2021sum,Freedman:2021ahq} for the TRGB, and \cite{Abdalla:2022yfr,DiValentino:2021izs,Verde:2019ivm} and references therein for an overview) and that indirectly inferred from the measurement of Cosmic Microwave Background (CMB) anisotropies (e.g., Planck \cite{Planck:2018vyg} or ACT, SPT \cite{ACT:2020gnv,SPT-3G:2021eoc}). This tension has been growing steadily for almost a decade \cite{Freedman:2017yms,Planck:2013pxb,Riess:2016jrr,Bernal:2016gxb,Verde:2013wza} and has consequently sparked a great amount of theoretical and experimental investigations. After the uninterrupted chain of  spectacular successes of the (standard) $\Lambda$CDM model over the past few decades, the high statistical significance of the tension has been interpreted as possible signature of new physics. While the theoretical side has focused on a wide variety of extensions of $\Lambda$CDM mainly influencing the CMB inference of $H_0$ (see \cite{DiValentino:2021izs,Schoneberg:2021qvd} among others), only a few models show promise. In parallel, on the experimental side a number of direct and indirect methods have been put forth to give alternative perspectives on the Hubble tension \cite{DiValentino:2021izs,Perivolaropoulos:2021jda}.

Given the far-reaching implications of a claim of new physics beyond $\Lambda$CDM, it is of utmost importance to verify that the tension does not arise from a single experiment, observation or data set. The physics of the early Universe is believed to be simple and well understood and the early-Universe probe {\it par excellence} is the CMB: the exquisite agreement of the $\Lambda$CDM model with CMB data represent one of the major successes of the model. However CMB constraints on $H_0$ are indirect and thus very model-dependent.

One particularly interesting combination of probes comes from the measurements of baryonic acoustic oscillations (BAO) within tracers of the dark matter (such as luminous red galaxies (LRG), emission line galaxies (ELG), quasi-stellar objects/quasars (QSO), and the Lyman-$\alpha$ forest (Ly$\alpha$)) combined with the determination of light element abundances that have been generated in the big bang nucleosynthesis (BBN). It is both independent of the CMB anisotropies altogether and perhaps the most tightly constraining combination of alternative data sets in terms of the Hubble tension.
In particular, one working hypothesis that has gained popularity is the early-vs-late split \cite{Bernal:2016gxb,Aylor:2018drw,Knox:2019rjx,Verde:2019ivm}, which can now be independently cross-checked. It is particularly interesting to notice that the early time solutions favored in \cite{Schoneberg:2021uak} can be further constrained independently of the CMB using the combination of BAO and BBN.

The great advantage is that it relies almost exclusively on quantities that can be determined at the background level in cosmological perturbation theory, thus being insensitive to some of the specifics of the dark matter and dark energy models. Since the BAO+BBN probes combine physics from far before (BBN) and far after (BAO) the time of recombination, they can provide an important cross-check of the assumed underlying model.

This combination is almost a decade old \cite{Addison:2013haa, Aubourg:2014yra} but the incredible improvement in the precision of the BAO measurements with the SDSS survey (especially DR12 in 2015) has re-sparked the interest in this combination a few years ago \cite{Addison:2017fdm,Blomqvist:2019rah,Cuceu:2019for}. Part of this renewed interest can also be attributed to the increased statistical power of  and  improved internal consistency (see \cite{Cuceu:2019for}) between Ly$\alpha$- and galaxy-based BAO measurements from the recent Ly$\alpha$ BAO measurement performed in the context of the eBOSS survey in \cite{Blomqvist:2019rah,deSainteAgathe:2019voe} compared to their DR11 counterparts \cite{BOSS:2013igd,BOSS:2014hwf}.

As opposed to the BBN-inspired priors on the baryon density in the aforementioned analyses, in \cite{Schoneberg:2019wmt} an analysis has been performed with a more detailed treatment of the BBN using a variety of different BBN codes (such as \cite{Pisanti:2007hk,Consiglio:2017pot,Pitrou:2018cgg}) and measurements (\cite{Cooke:2017cwo,Aver:2015iza,Peimbert:2016bdg,Izotov:2014fga}). Yet, only shortly thereafter, a new measurement of the BBN-energy deuterium burning rate by the LUNA experiment \cite{Mossa:2020gjc} has allowed for increased precision of the underlying BBN codes \cite{Gariazzo:2021iiu,Pitrou:2019nub}. Additionally, a new release of the BAO data from DR16 \cite{eBOSS:2020yzd,SDSS-IV:2019txh} has further decreased the uncertainties of the BAO measurements.

In this work we aim to extend the BAO+BBN analysis performed in \cite{Schoneberg:2019wmt} by using these current state-of-the-art data sets for both BBN and BAO, as well as furthering the investigation in several unexplored directions. 

\pagebreak[40]
In particular, we aim to discuss if previously known internal \enquote{tensions} still persist, to add robust alternative data from the measurement of clustering, to use the tight geometrical inference of the sound horizon angle from the CMB as an external prior, to add data from supernovae or cosmic chronometers, and to check the model-dependence of this probe with particular focus on models that could hinder the underlying mechanism behind this probe.

We describe the underlying physical mechanism behind the various probes as well as the corresponding data sets employed within this work in \cref{sec:method_and_data}. In \cref{sec:results} we present the results of our investigation and conclude in \cref{sec:conclusion}.

\section{Method and Data}\label{sec:method_and_data}
In \cref{ssec:method} we discuss the physical observables and corresponding constrained parameters from the various cosmological probes in greater detail, largely as a review of the physical mechanisms, though formulated in a condensed way with a slightly different perspective. Still, the expert reader might want to continue directly to \cref{ssec:data} where the data used within this work are described, or to \cref{sec:results} where the final results are shown.

\subsection{Method}\label{ssec:method}
Here we explain the physical mechanisms behind the various measurements employed within this work with particular focus on how their combinations might create synergies.

\enlargethispage*{4\baselineskip}
\subsubsection{The BAO+BBN probe}
The combination of BAO and BBN might seem a surprising one to strongly constrain $H_0$\,, as both probe physics from vastly different redshifts and neither of them is directly related to the Hubble parameter.
However, as we will explain below, it is exactly the complementary time-scales involved that allows for a CMB-independent view of the Hubble tension that is as constraining as the local distance ladder, and almost as constraining as CMB data.

The primary component of the BAO+BBN probe is the compilation of observations of the BAO in the clustering of galaxies, quasars, or in the intergalactic medium (IGM) as probed by the Lyman-$\alpha$ forest. This clustering carries the imprint of a standard ruler scale, the sound horizon at radiation drag $r_s$\,. Importantly, this standard ruler is not observed directly; instead, only its angular or redshift extent can be measured, leading to constraints on the combinations of $\delta z \approx H(z) r_s$ or $\delta \theta \approx r_s/r_A$.\footnote{Here $r_A$ is the usual co-moving angular diameter distance, i.e. $r_A = \int_0^{z} 1/H(x)\mathrm{d}x$ in a flat $\Lambda$CDM model. For the sound horizon $r_s$ see \cref{eq:soundhorizon}.} These two can be written in terms of the normalized expansion rate $E(z) = H(z)/H_0$ as $\delta z \approx (H_0 r_s) E(z)$ and $\delta \theta \approx (H_0 r_s) / \int [1/E(z)] \mathrm{d}z$. Notice that if $\delta z$ and $\delta \theta$ are measured for multiple redshifts, they can be used to disentangle the impact of the normalized expansion $E(z)$ and the product $H_0 r_s$\,. Since $E(z)$ contains only late-time density fractions (in flat $\Lambda$CDM it is entirely described by $\Omega_m$) it cannot be used for direct inference of the Hubble parameter. However, the simultaneously determined product of $H_0 r_s$ would allow for a direct measurement of $H_0$ if there was some kind of determination of the sound horizon $r_s$\, \cite{Bernal:2016gxb,Knox:2019rjx,Aylor:2018drw,Jedamzik:2020krr}. 

This is where the importance of the BBN becomes clear, since it allows for a calibration of $r_s$ using the measurement of the late-time normalized expansion rate. As further described in \cite{Schoneberg:2019wmt} the primordial abundances of light elements are very relevant both to the ratio of baryons to photons (effectively determining $\Omega_b h^2$ for a fixed\footnote{Effects of varying $T_\mathrm{CMB}$ are discussed in great detail in \cite{Ivanov:2020mfr}, but since the COBE/FIRAS experiment provides such an excellent measurement of $T_\mathrm{CMB}$ together with many other experiments, we consider it fixed to $2.72555\mathrm{K}$ for the remainder of this work.} $T_\mathrm{CMB}$) as well as the early time expansion rate (determined through any additional relativistic dark species at BBN for a fixed $T_\mathrm{CMB}$). The former is more directly constrained through the measurement of the deuterium abundance $D_H$ such as from \cite{Cooke:2017cwo}, while the latter is more constrained through the Helium abundance $Y_p$ (such as measured by \cite{Izotov:2014fga,Aver:2015iza,Peimbert:2016bdg}), though we stress that primarily their joint constraint is important 
(see also Fig. 1 of \cite{Schoneberg:2019wmt}). 

The BBN inference of $\Omega_b h^2$ in this case is what allows for the calibration of the sound horizon mentioned above. 
In fact, the sound horizon can be written as an integral
\begin{equation}\label{eq:soundhorizon}
    r_s = \int_{z_*}^\infty \frac{c_s(z)}{H(z)} \mathrm{d}z~.
\end{equation}
Na\"ively one could imagine that here too only the product $H_0 r_s$ could be determined for a given normalized expansion rate $E(z)$, leading to a perfect degeneracy. However, in this case the \emph{early Universe} normalized expansion rate does indirectly depend on $h$. This is because in $\Lambda$CDM-like and most popular extensions of this model, at high redshifts we have:
\begin{equation}\label{eq:highredshift_expansion}
    E(z) \approx \sqrt{\Omega_m (1+z)^3 + \omega_r/h^2 (1+z)^4},
\end{equation}
where the radiation density $\omega_r$  is fixed to
\begin{equation}\label{eq:radiationdensity}
    \omega_r \approx 2.47 \cdot 10^{-5} \left[1+\frac{7}{8}\left(\frac{4}{11}\right)^{4/3} N_\mathrm{eff}\right] \left(\frac{T_\mathrm{cmb}}{2.7255\mathrm{K}}\right)^4,
\end{equation}
which is measured independently of $h$, meaning that  \cref{eq:highredshift_expansion} does depend on $h$. We immediately notice a strong geometrical degeneracy between $h$ and $N_\mathrm{eff}$ at this background level. In principle, also the redshift of recombination depends on $h$, but this effect is far subdominant \cite{Schoneberg:2021uak}. All in all, for the cosmologies close to the Planck bestfit, the combination $H_0 r_s$ scales like\footnote{Instead of from numerical calculations like in \cite{Schoneberg:2021uak}, the same scaling could also be achieved when considering Eq. (26) of \cite{Eisenstein:1997ik} and powerlaw-expanding in the vicinity of the Planck bestfit, giving in $\Lambda$CDM $H_0 r_s \propto \Omega_m^{-0.23} h^{0.54} (\Omega_b h^2)^{-0.13}$, which is within the accuracy of the numerical formula. The numerical formula can also be checked easily for extended cosmologies (for example, one can easily find a $m_e^{-0.62}$ scaling relevant for \cref{ssec:me}).} \cite{Schoneberg:2021uak}
\begin{equation}\label{eq:rs}
    H_0 r_s \propto \Omega_m^{-0.23} h^{+0.52} N_\mathrm{eff}^{-0.1} (\Omega_b h^2)^{-0.11}.
\end{equation}
\enlargethispage*{2\baselineskip}%
The determination of $\Omega_b h^2$ (and possibly $N_\mathrm{eff}$ when it is not equal to 3.044) through BBN and the independent determination of $\Omega_m$ from BAO then allow $h$ to be uniquely determined from $H_0 r_s$\,. However, if the sound horizon $r_s$ is modified through an effect other than through \cref{eq:rs}, then BBN is not sufficient to calibrate it anymore. Naturally, this immediately implies that the BAO+BBN probe loses its constraining power. This is investigated in more depth in \cref{ssec:neff,ssec:me}.

\subsubsection{Additional probes}
Within this work, we will also be making use of unanchored Supernovae of Type Ia from Pantheon\cite{Pan-STARRS1:2017jku}/PantheonPLUS\cite{Scolnic:2021amr,Brout:2022vxf} as well as Cosmic Chronometer data from \cite{Moresco:2018xdr,Moresco:2020fbm}.

The supernovae of type Ia allow for a measurement of the normalized expansion rate $E(z)$ since the ratio of the observed and intrinsic\footnote{Naturally these intrinsic luminosities are not measured, but determined through the standardization procedure, by using the parameters of the light curve to estimate this quantity.} luminosities directly determines the product $H_0 D_L(z) = (1+z)\int_0^z [1/E(x)] \mathrm{d}x$\,. This, in turn allows for a direct determination of $\Omega_m$ for a flat $\Lambda$CDM model, which can aid in the determination of the sound horizon $r_s$\,.

The Cosmic Chronometers (CC) method relies on determining the differential age $\delta t$ and the differential redshifts $\delta z$ of stellar populations within old massive galaxies that have not undergone strong recent star formation. See the excellent review by \cite{Moresco:2022phi} for more details. Effectively, CC can thus directly measure the Hubble rate $H(z) \approx -\delta z/\delta t /(1+z)$ and thus, for a given measured normalized expansion rate  $E(z)$ also determine the Hubble parameter. These CC measurements are a crucial cross-check of other methods,
but -- without some additional assumptions -- do not yet allow for such a tightly constrained determination of either $\Omega_m$ or $h$ alone, which would be needed for a decisive statement about the Hubble tension. Even when combined with other un-anchored late-universe probes, the determination of $h$ is limited by the CC probe -- as all other un-anchored probes currently available measure only $E(z)$, not $H(z)$ directly.

\subsubsection{ShapeFit}
The addition of redshift space distortions (RSD) and in particular the ShapeFit (SF) results are an interesting and novel aspect of this work. While the RSD probe determining the parameter combination $f \sigma_8$ is well known within the literature \cite{eBOSS:2020yzd}, the SF approach is comparatively new. We refer the reader to \cite{Brieden:2021edu,Brieden:2021cfg,Brieden:2022lsd} for a more complete review, and only summarize here the basic features relevant to the BAO+BBN probe. The SF is based on measuring the effective slope of the power spectrum at an intermediate scale of around $0.03h/\mathrm{Mpc}$. This slope is influenced by mainly three effects.
\begin{enumerate}
    \item The baryon suppression, which is caused by the reduced growth at scales roughly larger than $1/r_s \sim 0.01h/\mathrm{Mpc}$ due to the acoustic oscillations of the baryons preventing their infall in gravitational wells. The amplitude of this suppression is determined through the ratio $[\Omega_b h^2]/[\Omega_m h^2]$\,.
    \item The turnover point of the power spectrum, given by the equality scale $k_\mathrm{eq}$, which is determined through the redshift of equality $1+z_\mathrm{eq} = [\Omega_m h^2]/\omega_r$\,.
    \item The overall tilt of the power spectrum, which is directly related to the primordial power spectrum tilt $n_s$\,.
\end{enumerate}
Thus, we can see that if SF is calibrated through $n_s$ (from a prior) as well as supplemented through a measurement of $\{\Omega_b h^2, \Omega_m\}$ through the BAO+BBN combination, it can indeed provide an additional measurement of $h$. This additional constraining power arises from the effects on the matter transfer function of processes around and before the time of matter radiation equality (baryon suppression and turnover).

This method attempts to isolate the information coming from the large, linear scales as compared to the FullModeling analysis\footnote{By this we mean an analysis that uses the full shape of the measured power spectrum for direct inferences on a given model, as opposed to measuring only particular features of it, such as the BAO with $\{\alpha_\parallel, \alpha_\perp \}$ or the RSD with $\{f \sigma_8\}$.} (see for example \cite{Ivanov:2019pdj,DAmico:2019fhj,Philcox:2020vvt,Chudaykin:2020aoj,Wadekar:2020hax,Kobayashi:2021oud,Chen:2021wdi}), which relies on perturbative expansions of the power spectrum in order to perform the full modeling of the power spectrum shape. Due to measurements of the broadband shape of the power spectrum, these methods can independently infer $H_0$ with a relatively high precision (though often they also rely on fixing the baryon density and the primordial spectral index). Particularly, \cite{Smith:2022iax} recently showed that by marginalizing over $r_s$ these constraints can be obtained for the Hubble parameter fully independently of the sound horizon information (which is the basis of the BAO+BBN probe) and apply even to extended cosmologies. We will leave a more thorough comparison to FullModeling analyses for future work, but mention that the SF approach has been shown in \cite{Brieden:2021cfg} to provide virtually the same constraining power as FullModeling analyses if  similar priors on the primordial spectral index and the baryon density are adopted. The latter, for the purpose of this work, will of course not be necessary and instead will be replaced by the full BBN likelihood (which is further described in \cite{Schoneberg:2019wmt}).

\subsection{Data}\label{ssec:data}
We make use of several recent advancements in the measurement of the BAO as well as more precise BBN theoretical predictions from updated interaction rates from the LUNA experiment. We list the data sets we use as well as a short acronym for each below.

\begin{itemize}
    \item \textbf{BAO} We use the newest release of the reconstructed BAO parameters ($\alpha_\parallel\,,\alpha_\perp)$ from DR16 \cite{eBOSS:2020yzd,Bautista:2020ahg,Gil-Marin:2020bct,Neveux:2020voa,Hou:2020rse}, including LRG and QSO. Additionally, this includes high redshift BAO data from the Lyman-$\alpha$ forest\footnote{When this article was at a relatively advanced stage, the Lyman-$\alpha$ forest data from DR16 became public. The difference between DR16 and DR14  Lyman-$\alpha$ forest data is unimportant for the findings of this work (see also \cref{ssec:internaltension}).} being  measured in the DR14 \cite{Blomqvist:2019rah,deSainteAgathe:2019voe}. We also provide a version \textbf{BAO (DR12)} that includes older DR12 \cite{BOSS:2016wmc} data instead of the DR16 data. In this combination, the Lyman-$\alpha$ forest data is unchanged.
    \item \textbf{RSD} We use the redshift distortion data measuring the data combination $f \sigma_8$ from the clustering in the newest DR16 data (LRG and QSO) \cite{Gil-Marin:2020bct,Hou:2020rse,Bautista:2020ahg,Neveux:2020voa}.
    \item \textbf{SF} We use the measurement of the ShapeFit parameter $m$ from \cite{Brieden:2022lsd}. By default, we also use a prior in $n_s = 0.9637 \pm 0.0044$ motivated from \cite{Planck:2018vyg} as in the original work.
    \item \textbf{BBN} We use the Big Bang Nucleosynthesis constraints from Aver~et~al.~\cite{Aver:2015iza} for Helium and Cooke~et~al.~\cite{Cooke:2017cwo} for Deuterium, using newest theoretical predictions from \texttt{Parthenope 3.0} incorporating the LUNA results \cite{Mossa:2020gjc} on the Deuterium-proton to Helium3-photon fusion cross section. We also provide a version \textbf{BBN(noLUNA)} that does not include the new theoretical predictions based on the LUNA results, but instead uses older predictions from \texttt{Parthenope 2.0} \cite{Consiglio:2017pot}. \footnote{In both cases we add a theoretical uncertainty of $6 \cdot 10^{-7}$ for the deuterium abundance, in accordance with \cite{Gariazzo:2021iiu} for the new rates and approximately with \cite{Consiglio:2017pot} for the old rates. This theoretical uncertainty on deuterium is in decent agreement with the systematic uncertainties from \cite{Yeh:2022heq}. We also put a $0.0003$ theoretical uncertainty for the Helium abundance (as in both \cite{Consiglio:2017pot,Gariazzo:2021iiu}), which is propagated from the uncertainty of the neutron lifetime.}
    \item $\boldsymbol{\theta_s}$ We impose a prior ($\theta_s = 1.04109 \pm 0.00030$) on the sound horizon angle as measured by the CMB \cite{Planck:2018vyg}. This quantity can also be seen as the result of a (massive) data compression process, encoding the angular position of the CMB acoustic peaks; as such  this determination is almost entirely model-independent. The model dependence comes in almost exclusively at the interpretation stage.
    \item \textbf{PantheonPLUS} \cite{Brout:2022vxf} We use the newest compilation of uncalibrated supernovae of Type Ia, constraining $\Omega_m = 0.338 \pm 0.018$ in a flat $\Lambda$CDM model. We also provide a version with the older \textbf{Pantheon} data \cite{Pan-STARRS1:2017jku}, which constrains $\Omega_m = 0.298 \pm 0.022$ in a flat $\Lambda$CDM model. We use the full likelihoods (not $\Omega_m$ priors) in both cases.
    \item \textbf{CC} We use Cosmic Chronometers data  from \cite{Moresco:2020fbm}, which constrain $H(z)$ by measuring differential ages and redshifts of passive stellar populations within mostly early massive galaxies. We follow the prescription by \cite{Moresco:2020fbm} to include the off-diagonal covariance matrix entries.
\end{itemize}

For each data combination we run a \texttt{MontePython} MCMC chain in order to obtain the corresponding constraints, using \texttt{class} as the underlying cosmological code. Unless otherwise stated we require a Gelman-Rubin convergence criterion of $|R-1| < 0.01$ for all parameters to ensure proper convergence of all chains.

\section{Results}\label{sec:results}
In \cref{ssec:newdata} we update the BAO+BBN analysis of \cite{Schoneberg:2019wmt} to include the new developments in data and modeling discussed in \cref{ssec:data}. In \cref{ssec:internaltension} we discuss the contribution to the final constraints coming from different redshifts, and continue in \cref{ssec:shapefit} to investigate the impact of the new ShapeFit results. In \cref{ssec:thetas} we investigate the addition of a geometric sound horizon angle prior from the CMB and in \cref{ssec:pantheon,ssec:cc} we continue by adding supernovae and cosmic chronometer data. These results of this section initially assume a $\Lambda$CDM cosmology, though in \cref{ssec:neff,ssec:me} this $\Lambda$CDM assumption will be relaxed in order to check the model-dependence of the BAO+BBN combination. All results are also summarized in \cref{tab:results}.

\subsection{Impact of updated data}\label{ssec:newdata}

\begin{figure}[t]
    \centering
    \includegraphics[width=0.63\textwidth]{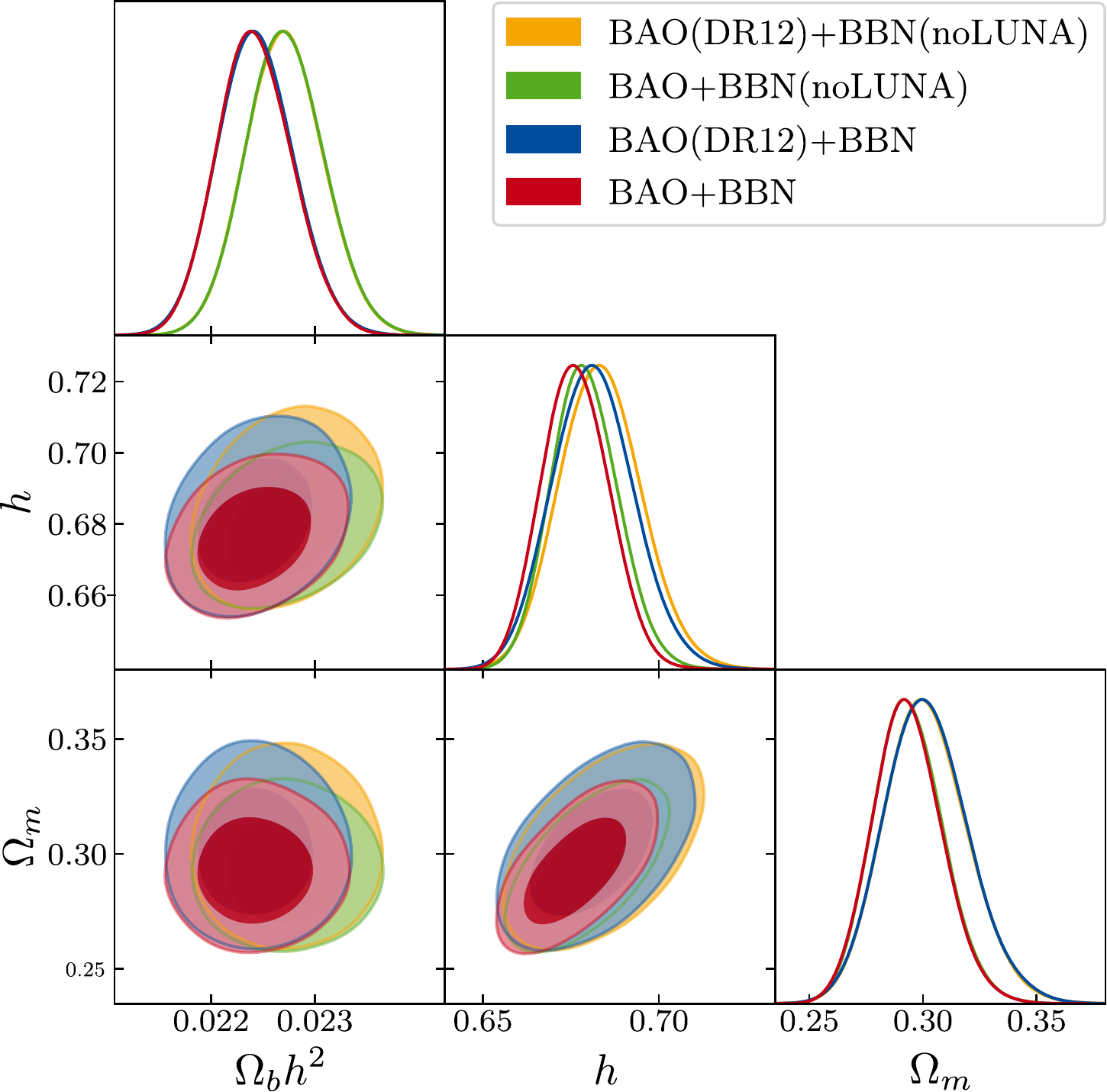}
    \caption{Impact on key parameters in the $\Lambda$CDM model of updating the BAO LRG data from DR12 to DR16 and updating the BBN predictions from \texttt{Parthenope~2.0} (noLUNA) to LUNA-based theoretical predictions from \texttt{Parthenope~3.0}.}
    \label{fig:data}
\end{figure}

We show the impact of updating both the BAO and BBN data on the cosmological parameters constraints in \cref{fig:data}. 

To quantify the shifts in parameters as well as the change in the uncertainty to a high degree of precision, we require instead $|R-1| < 10^{-4}$ for these MCMC chains.

There is a shift in $\Omega_b h^2$ from the updated BBN data, as the new theoretical predictions of BBN from \texttt{Parthenope 3.0} incorporating the new LUNA results yield a decreased deuterium abundance for a given $\Omega_b h^2$\,. Due to their anti-correlation, this results in lower values of $\Omega_b h^2$ being preferred in order to give the same measured deuterium abundance. The shift is only at a level of $0.8\sigma$ ($\Omega_b h^2 = 0.02272 \pm 0.00038$ with the old predictions vs $\Omega_b h^2 = 0.02242 \pm 0.00037$ with the new LUNA-based predictions), and more importantly does not result in a significant shift of $\Omega_m$ ($0.02\sigma$). The shift in $H_0$ ($0.3\sigma$) due to the re-calibration of the sound horizon is also rather minor, showing the small dependence of the results on the precise assumed value of $\Omega_b h^2$ from BBN.

\enlargethispage*{1\baselineskip}
Not unexpectedly (given the larger volume surveyed), the new BAO data from DR16 do shift the constraint in $\Omega_m$ by around $0.4\sigma$ ($\Omega_m = 0.30171 \pm 0.018744$ with DR12 data vs $\Omega_m = 0.29338 \pm 0.015434$ with DR16 data), resulting in a direct shift of $H_0$ by $0.4\sigma$. We also note that due to the tightening of the constraint on $\Omega_m$ by about $20\%$, the constraint on $H_0$ is similarly tightened by about $20\%$.

Taken together, these two effects modify the determination of the Hubble parameter from $H_0 = \cons{68.3}{1.1}{1.2}$
km/s/Mpc for the old analysis to
$H_0 = \cons{67.6}{0.9}{1.0}$
km/s/Mpc for the updated analysis, a shift of the constraint of about $0.64 \sigma$ and a tightening of its uncertainty of  about $20\%$, leading to a total tension of around $3.7 \sigma$ with the newest Pantheon+SH0ES results from \cite{Riess:2021jrx} (up from $3.2\sigma$ in \cite{Schoneberg:2019wmt}).

\subsection{Impact of internal (in)consistencies}\label{ssec:internaltension}

It is well known that the different dependence of $r_A$ on $\{\Omega_m, h r_s\}$ for different redshifts is crucial in aiding the BAO to achieve a tight constraint on $\Omega_m$ (as otherwise $\Omega_m$ and $h r_s$ would be somewhat degenerate). In \cite{Schoneberg:2019wmt} we showed that the high- and low-redshift data had quite striking different redshift dependencies (see also \cite{Schoneberg:2021uak}), and their intersection happened to lie at a low $h$ value (see Fig. 2 of \cite{Schoneberg:2019wmt}). At times it has been argued that the low $h$ value preferred by this combination was primarily driven by the internal inconsistency (sometimes even dubbed \enquote{tension}, see for example \cite{Cuceu:2019for}) between the high and low-redshift data sets. This concern was also driven by fears that the Ly$\alpha$ BAO could be impacted by systematics in the modeling of the intergalacitc medium (IGM). 
While \cite{duMasdesBourboux:2020pck} shows a great level of consistency across different analyses and data releases,
we want to investigate the dependence of our constraint on the high-redshift data and the Ly$\alpha$ BAO in particular.

\begin{figure}[H]
    \centering
    \includegraphics[width=0.45\textwidth]{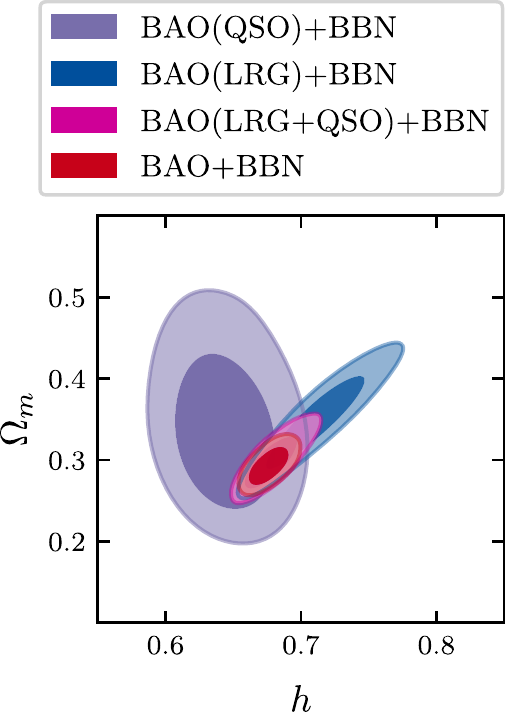}
    \includegraphics[width=0.5\textwidth]{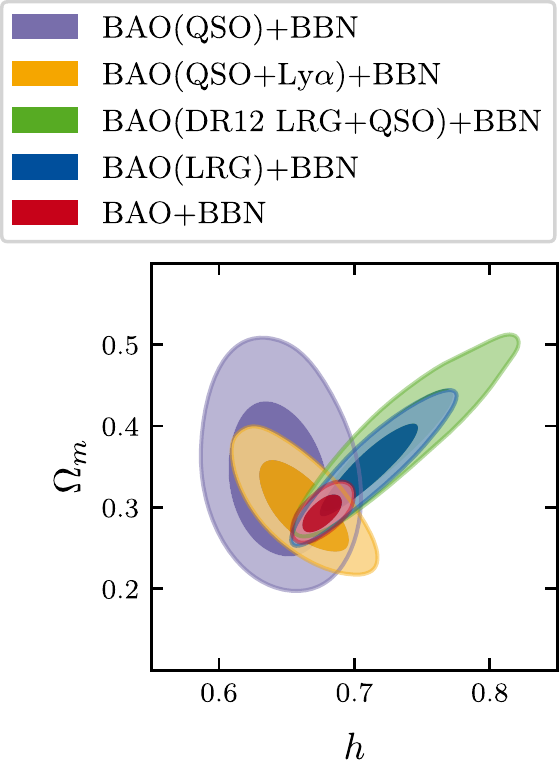}
    \caption{Constraints (in a $\Lambda$CDM model) on selected parameters from different subsets of the DR16 BAO data separated by their type -- Lyman-$\alpha$ forest based (Ly$\alpha$), quasar-based (QSO), or based on luminous red galaxies (LRG) compared to the full BAO+BBN probe (LRG+QSO+Ly$\alpha$). \textbf{Left:} Variations with/without QSO, always without Ly$\alpha$. \textbf{Right:} Variations from including Ly$\alpha$ or older DR12 BAO data.}
    \label{fig:redshiftsplit}
\end{figure}

In \cref{fig:redshiftsplit} we show the resulting constraints from combining BBN data with various subsets of the BAO data.\footnote{Compare also to fig 5 (right panel) of \cite{eBOSS:2020yzd}. We include a few additional sub-divisions of the data, in order to track the differences compared to earlier BAO+BBN combinations.}

The results are \emph{largely independent of using BAO measurements from the Lyman-$\alpha$ forest}, as the constraints with and without Lyman-$\alpha$ based BAO almost perfectly overlap. This is different from \cite{Schoneberg:2019wmt} where the high redshift lever-arm of the Lyman-$\alpha$ data was crucial for obtaining the tight constraints. Now the long lever-arm in redshift offered by the QSO-based BAO mean that the $\Omega_m-h r_s$ (or, with a BBN prior, the $\Omega_m-h $ degeneracy) can be broken even without the information from the Lyman-$\alpha$ forest data (as indicated by the red and blue contours in the left panel of \cref{fig:redshiftsplit}).

The 1$\sigma$ contours of the high-redshift data, both from the QSO alone and from QSO together with the Ly$\alpha$, now overlap with those of the LRG at lower redshifts (see the dark blue, green, and red contours in \cref{fig:redshiftsplit}). This means that \emph{it can no longer be argued that the joint constraint relies on an internal tension} to artificially strengthen the $\Omega_m$ constraint (and thus enforce a stricter bound on $h$).

Note that the data based on the Lyman-$\alpha$ forest do significantly help the QSO determination in the $\Omega_m-h$ degeneracy plane (right panel of \cref{fig:redshiftsplit}, green vs dark blue contours), and are perfectly in agreement with the region preferred by the LRG data. This demonstrates again the multiple layers of consistency present in the BAO data, and in particular that \emph{the Lyman-$\alpha$ based data are perfectly compatible with the rest}.

Finally, part of the reason why the agreement also looks visually better is that the update of DR12 to DR16 LRG data did not only shrink the overall error bars but also slightly shifted the contour towards the region of lower $h$ and $\Omega_m$ along the degeneracy direction.

As a consequence, in the following sections we always include the Lyman-$\alpha$ and QSO-based high-redshift BAO data.

\subsection{Impact of clustering information beyond BAO}\label{ssec:shapefit}
The clustering of tracers of the dark matter can  be used  beyond the \enquote{minimalistic} goal to constrain the classical BAO scaling parameters $(\alpha_\parallel\,, \alpha_\perp)$ through the measurement of the BAO peak position. Depending on additional assumptions, one can either perform a FullModeling analysis or only include extra information from the RSD (see \cref{ssec:method} for terminology). 
In the recently-proposed ShapeFit (SF) approach the  parameter $m$ is added to the set of parameters in addition to the usual RSD combination $f \sigma_8$\,, leading to virtually the same constraining power as in the FullModeling analysis (though with increased robustness). 

\begin{figure}[H]
    \centering
    \includegraphics[width=0.75\textwidth]{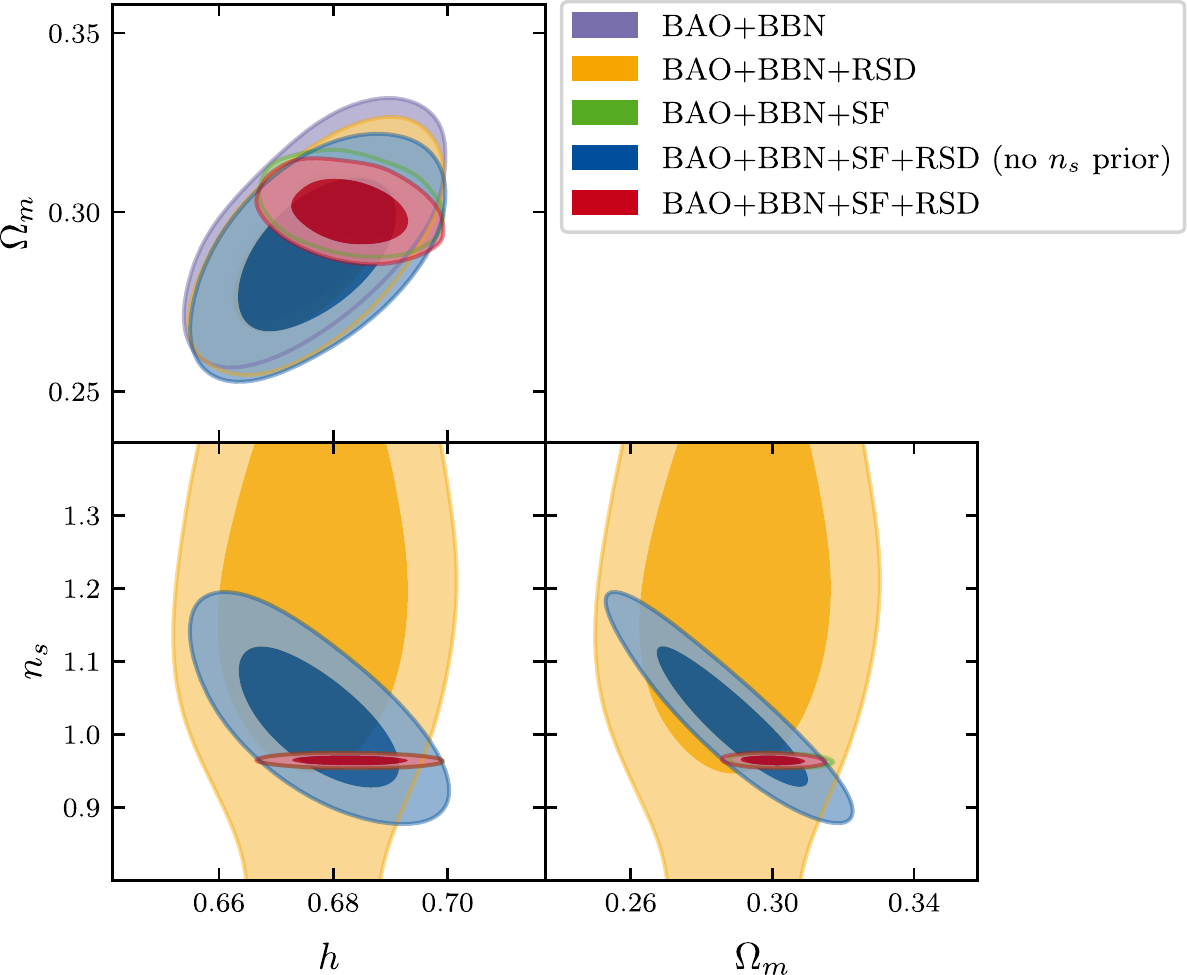}\\[2em]
    \includegraphics[width=0.75\textwidth]{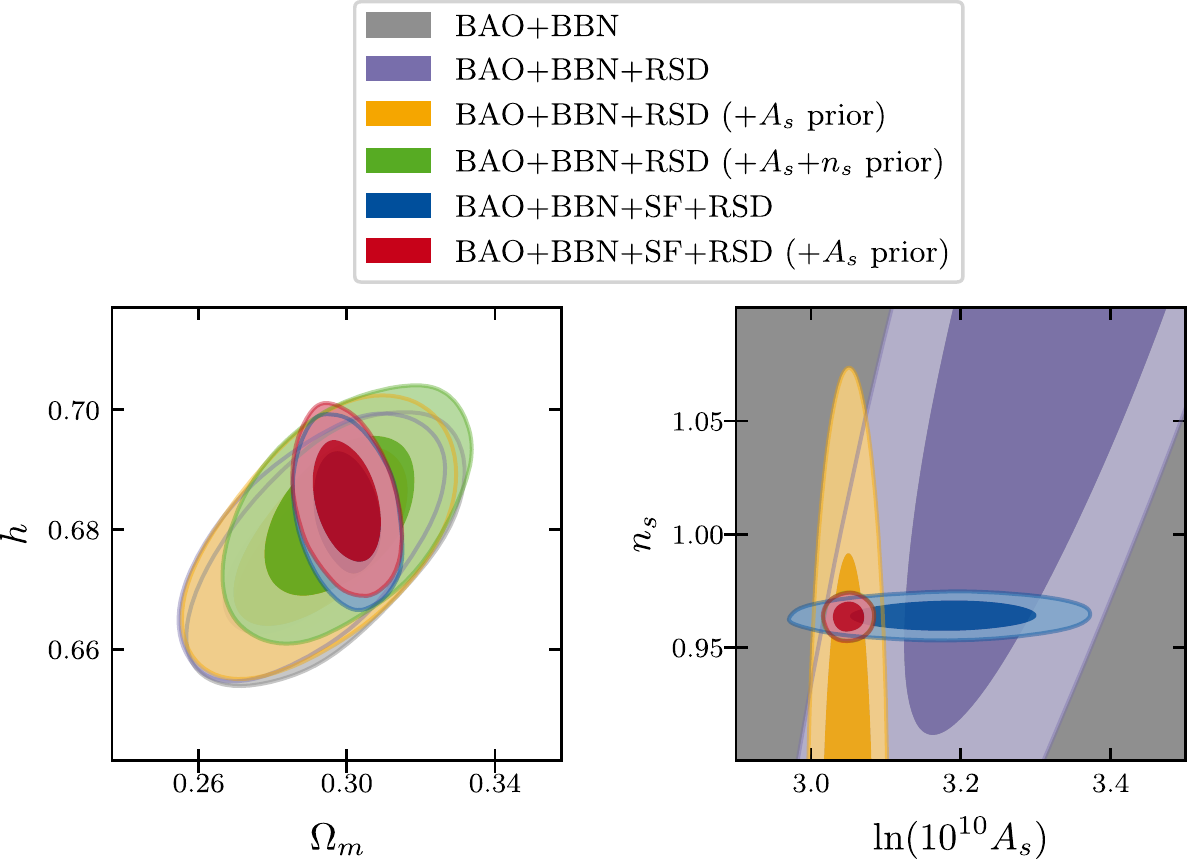}
    \caption{Impact on selected parameters for a $\Lambda$CDM model of the redshift space distortion (RSD) and ShapeFit (SF) data (see \cref{ssec:data}). Additionally, we impose priors on $n_s$ for the SF by default, and only optionally on $A_s$ and/or $n_s$ for the RSD. \textbf{Upper panel:} Combinations of SF with or without $n_s$ prior.  Note that the green contours almost perfectly overlap with the red contours, indicating that RSD does not add a lot of information once SF is included. \textbf{Lower panel:} Combinations of RSD with various priors on $A_s$ and/or $n_s$\,. Only when RSD is combined with both priors (green) are the constraints in $h$ somewhat competitive with those of SF (blue), otherwise overlap with those without RSD (grey/purple/orange).}
    \label{fig:shapefit}
\end{figure}

The constraints resulting from including RSD or SF analyses from DR16 LRG are shown in \cref{fig:shapefit}. In the top panel we compare the gain in constraining power from RSD and SF data, using the original prescription for both (no prior for RSD, a prior on $n_s$ for SF of  $n_s = 0.9637 \pm 0.0044$ motivated by \cite{Planck:2018vyg}). We see that while the gain in constraining power from RSD is rather minute, SF does significantly shrink the contours. The error bars on $h$ shrink by a factor of $1.3$, while the error bars on $\Omega_m$ shrink by a factor of $2.0$, leading to a measurement of $H_0 = \cons{68.29}{0.68}{0.69} \mathrm{km/s/Mpc}$ ($3.8 \sigma$ tension). The reason why SF is expected to give a good additional constraint in the $\{\Omega_m,h\}$ plane is explained in \cref{ssec:method}. If we instead drop the prior on $n_s$ for the SF measurement, it looses its constraining power, due to the strong $m-n_s$ degeneracy explained further in \cite{Brieden:2021edu} and in \cref{ssec:method}. We conclude that the SF (with the default $n_s$ prior) strongly supports the existing measurements, while SF without a $n_s$ prior provides no additional constraining power.

As such, we might ask if the RSD constraints would also benefit from a prior on~$A_s$\,, and this is investigated in the bottom panel of \cref{fig:shapefit}. We find that with or without a prior on~$A_s$ (here we use $\ln(10^{10}A_s) = 3.047 \pm 0.014$) the constraints on $\{H_0, \Omega_m\}$ are virtually the same. This is because of the strong remaining degeneracy of $\sigma_8$ (and thus $\Omega_m$ and $h$) with~$n_s$\,. Only when including  an additional prior on $n_s$ can the RSD provide a small amount of additional constraining power\footnote{Of course, this stems from the fact that RSD measure $f \sigma_8$, which, in the limit of a tightly constrained~$\sigma_8$\,, measures mostly $\Omega_m$\,. In turn, combined with the BAO+BBN probe, this  can be converted into a measurement of $h$.}. However, we also notice that the SF measurement is putting much tighter constraints in the $\{H_0, \Omega_m\}$ plane, and the addition of RSD even with the $A_s$ prior (and the default $n_s$ prior of SF) does not lead to a  further improvement of the constraints.
%

Summarizing, a prior on $A_s$ from Planck does not significantly increase the constraining power of the RSD (unlike the $n_s$ prior for SF). Only when including a prior both on $A_s$ and $n_s$ we find that the parameter $\sigma_8$ is constrained enough to obtain a competitive inference of $\Omega_m$ from RSD, leading to slightly tightened constraints on $h$ for the BAO+BBN combination. However, even this increase in constraining power is much smaller than that obtained from the addition of the SF data. While RSD thus remains an excellent probe of theories of gravity, we find that it is outperformed by the SF on its impact on the Hubble tension.

One can understand why RSD has such a comparatively weak performance\footnote{RSD is a very good at detecting mismatches between growth and expansion history, though, and thus is an excellent probe of extended models of gravity beyond general relativity.} compared to BAO and SF in this context as follows.  RSD gives constraints on $\Omega_m$ (through $f(z)$) which are superseded by the excellent precision of the $\Omega_m$ determination from the \AP{} effect of the BAO. This holds for any model leaving the late-time expansion history unchanged, as we confirm in \cref{ssec:neff}.

\subsection{Impact of the CMB sound horizon angle}\label{ssec:thetas}
\begin{figure}[H]
    \centering
    \includegraphics[width=0.8\textwidth]{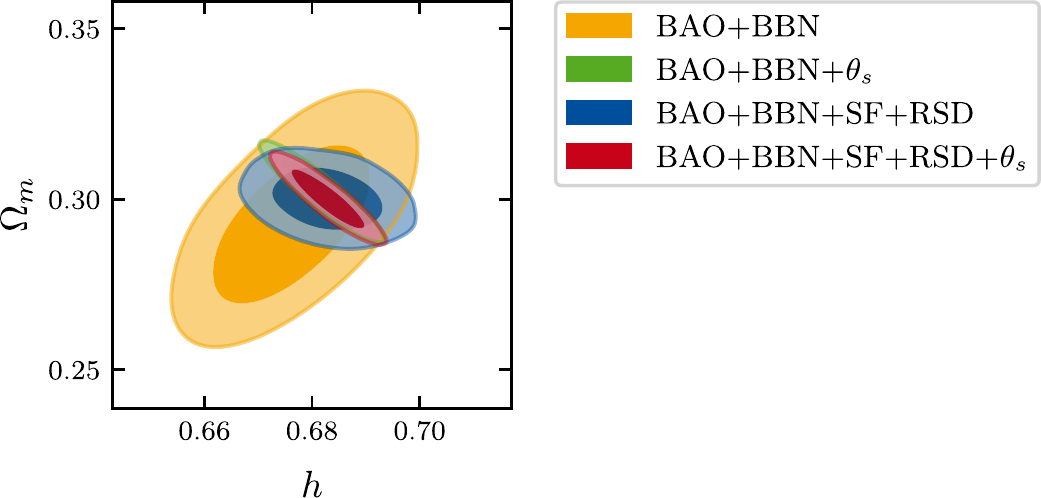}
    \caption{$\Lambda$CDM model parameters constraints for BAO+BBN when including a $\theta_s$ prior compared to those without. SF denotes ShapeFit, and RSD denotes redshift space distortions. The region preferred by BAO+BBN with a sound horizon prior (green) agrees very well with the region preferred by BAO+BBN with SF+RSD (blue).}
    \label{fig:thetas}
\end{figure}

The apparent angular position of the CMB peaks can be measured in an almost model-independent way. The CMB sound horizon angle, $\theta_s$\,, corresponds almost directly to the location of the various peaks in the angular correlation function, and is thus measured to be about the same value in $\Lambda$CDM as well as for models with strong modifications in early or late-time physics. As such, this rather geometric probe can be used as an interesting cross-check. We show the constraints with a prior on $\theta_s$ according to \cref{ssec:data} in \cref{fig:thetas}.

The predominant effect of the inclusion of the sound horizon angle information is a much tighter constraint on $\Omega_m$ and (through the BBN-calibration of $r_s$) correspondingly on $h$. This is to be expected since $\theta_s$ has a largely orthogonal dependence on $\Omega_m$ compared to the BAO shift parameters due to the large difference in redshift. The mechanism is thus not unlike the way that low and high redshift BAO combine to give a more precise measurement of $\Omega_m$ in the first place (see fig. 2 of \cite{Schoneberg:2019wmt}). In this case, the $\Omega_m$ constraint is increased by a factor of 2.5. For a BBN-calibrated $r_s$ this results in a factor of 2.0 for the constraints in $h$ to $H_0 = 68.16 \pm 0.45 \mathrm{km/s/Mpc}$ (without SF) and $H_0 = 68.30 \pm 0.45 \mathrm{km/s/Mpc}$ (with SF), leading to an overall tension of $4.2\sigma$ (with or without SF). The constraints of the updated BAO+BBN probe together with the geometrical sound horizon information are now as tight as the original analysis of CMB+CMB lensing+BAO was in the original analysis.

\enlargethispage*{4\baselineskip}
We stress that the region of $\Omega_m$ preferred by the inclusion of this $\theta_s$ prior is the same region of $\Omega_m$ preferred by the ShapeFit parameter $m$ (with Planck priors on $n_s$), suggesting a rather high level of compatibility between these two approaches and consistency of how the model describes the relevant physical processes across widely different epochs in the history of the  Universe (BBN, matter-radiation equality, recombination, late-time evolution). Combining the ShapeFit and $\theta_s$ results does not lead to a much stronger increase in constraining power since the ShapeFit parameter is just as accurate as the $\theta_s$ slice in terms of $\Omega_m$\,. We can thus expect  that upcoming large-scale structure measurements will provide an important consistency check for the Hubble tension at this robust geometrical level.

\subsection{Impact of Pantheon data}\label{ssec:pantheon}

\begin{figure}[H]
    \centering
    \includegraphics[width=0.52\textwidth]{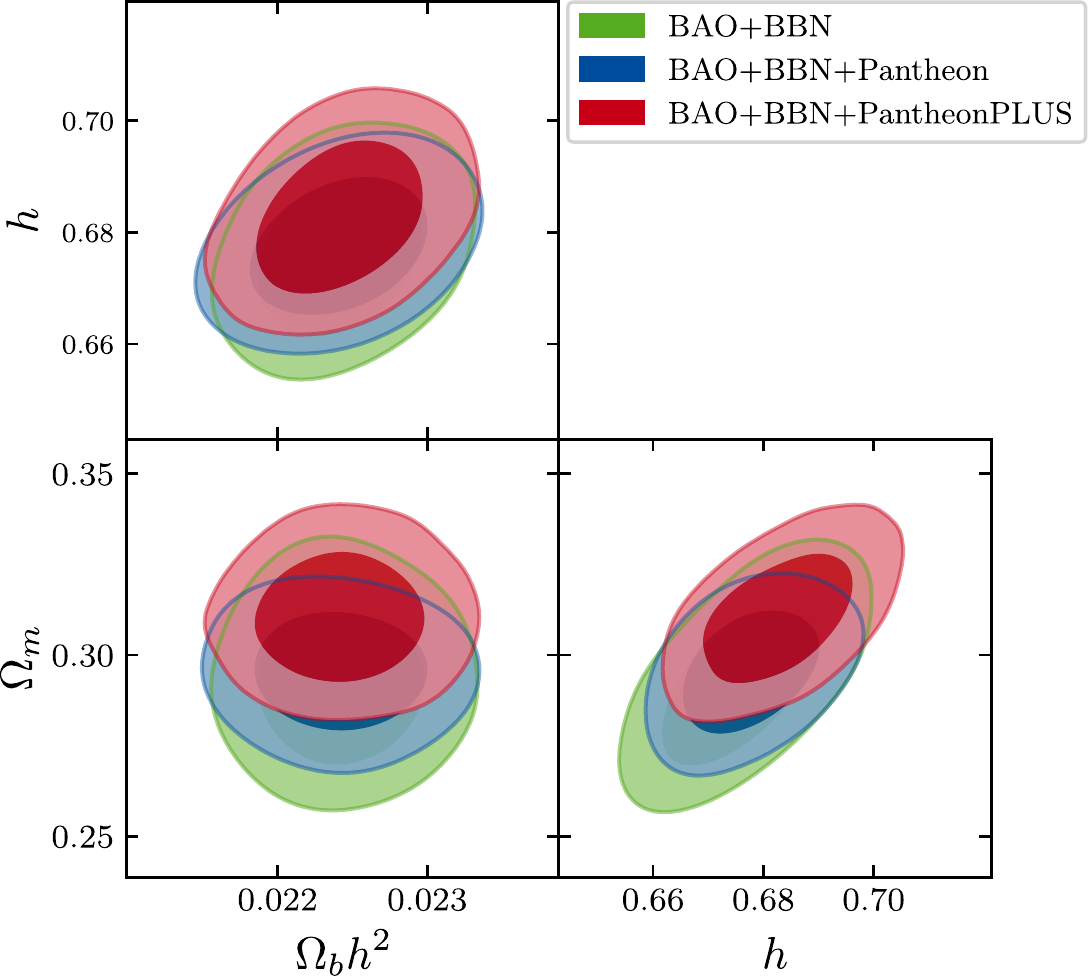}
    \includegraphics[width=0.47\textwidth]{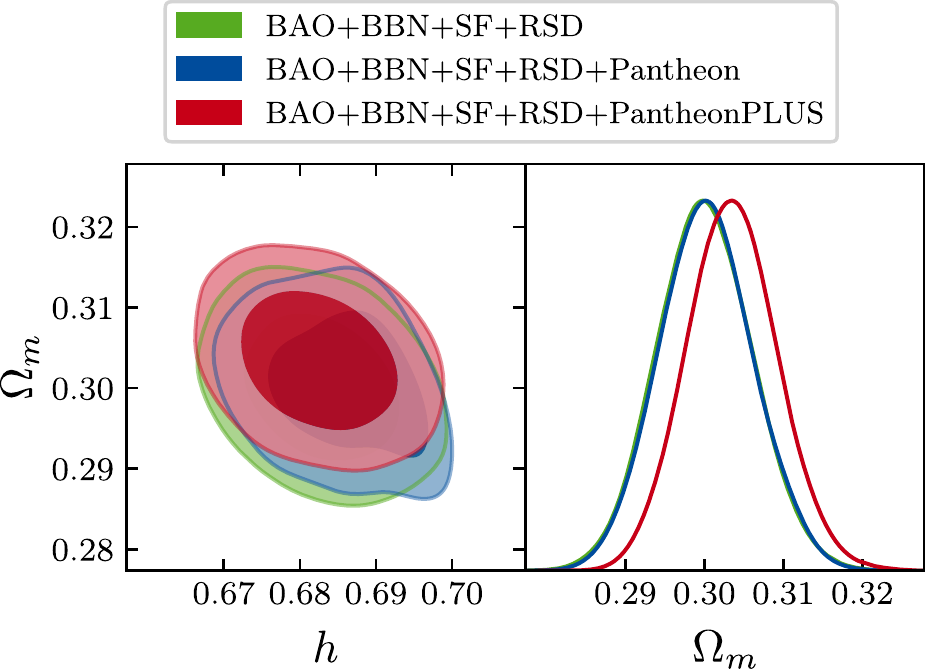}
    \caption{Impact of including old Pantheon or new Pantheon+ (PantheonPLUS) data on selected parameters constraints for a $\Lambda$CDM model. SF denotes ShapeFit, and RSD denotes redshift space distortions. \textbf{Left:} Without using any SF or RSD data. There is no significant change in $\Omega_b h^2$ between the data sets. \textbf{Right:} Including SF and RSD data. We also show the 1D posterior on $\Omega_m$\,, which shows a slight shift for PantheonPLUS compared to the analysis without it.}
    \label{fig:pantheon}
\end{figure}

Supernovae of type Ia are well known to provide strong constraints on the matter abundance $\Omega_m$ even in the absence of an absolute calibration. In \cref{fig:pantheon} we show the impact of including such uncalibrated supernovae using the Pantheon data set. We either show the impact of the updated Pantheon+ data (preferring a slightly higher value of $\Omega_m \approx 0.338 \pm 0.018$ \cite{Brout:2022vxf}) or the old Pantheon catalogue data (preferring a slightly lower value of $\Omega_m \approx 0.284 \pm 0.012$ \cite{Pan-STARRS1:2017jku}).
Since the $\Omega_m$ value preferred by the updated data set has changed slightly, we show show both data sets separately for better comparison with previous literature and for appreciating the effect of this change.

The additional $\Omega_m$ constraint from the old Pantheon data is in good agreement with the $\Omega_m$ preferred by the combination of BAO measurements, while the new PantheonPLUS data prefers a slightly (around $\sim1.9\sigma$) higher value. Without SF, the shift in the Hubble parameter caused by this shift in $\Omega_m$ is rather minor, leading to $H_0 = \cons{67.75}{0.70}{0.84} \mathrm{km/s/Mpc}$ (Pantheon,  $4.2\sigma$) or $H_0 = \cons{68.27}{0.85}{0.98} \mathrm{km/s/Mpc}$ (PantheonPLUS, $3.5\sigma$). The corresponding (perhaps unexpected) decrease in constraining power for PantheonPLUS compared to the old Pantheon data is naturally explained by the discrepancy in the singular $\Omega_m$ constraints with BAO in PantheonPLUS, leading to widened combined constraints (which propagate to widened constraints on $H_0 r_s$ and thus on $h$). If we instead include SF data, the results are remarkably stable and no significant shift in $H_0$ is observed.

\subsection{Impact of Cosmic Chronometer Data}\label{ssec:cc}

\begin{figure}[H]
    \centering
    \includegraphics[width=0.36\textwidth]{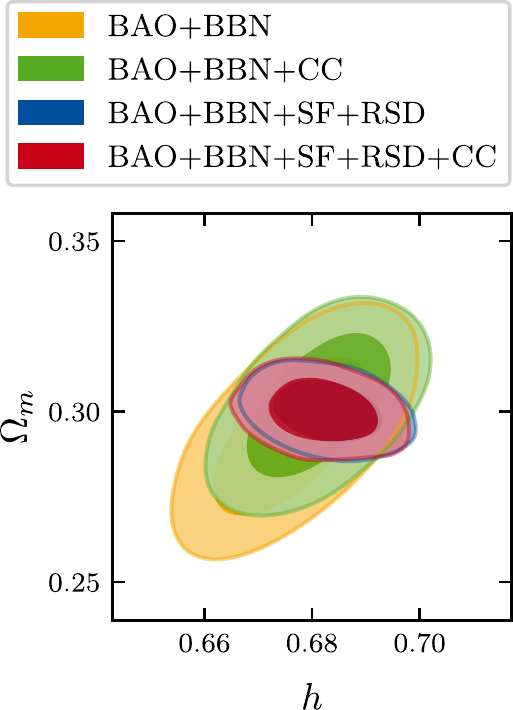}
    \includegraphics[width=0.52\textwidth]{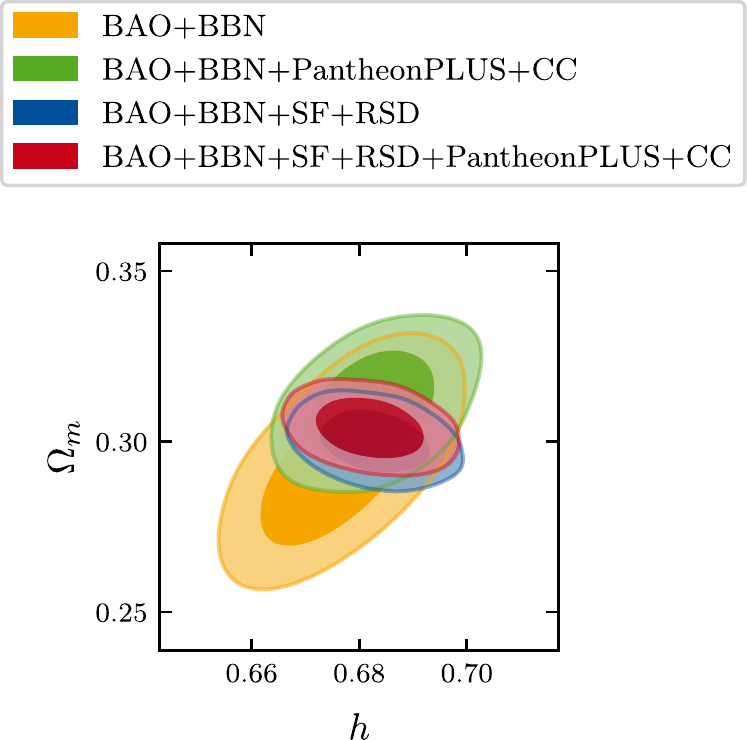}
    \caption{Impact of including cosmic chronometers (CC) data with and without PantheonPLUS data (always for a $\Lambda$CDM model). SF denotes ShapeFit, and RSD denotes redshift space distortions. \textbf{Left:} Without PantheonPLUS data. \textbf{Right:} With PantheonPLUS data.}
    \label{fig:cc_pantheon}
\end{figure}

The cosmic chronometers (CC) are an excellent late-universe probe for both $\Omega_m$ and more directly $H_0$ due to their direct measurement of $H(z)$. We show the corresponding results in \cref{fig:cc_pantheon}. We find that the addition of CC is mostly relevant when not including SF or RSD data, since, in this context, those supersede the constraining power of the CC. As a result, the CC can make the measurement more robust, but do not strengthen the constraint a lot by themselves. For the combination without SF and RSD we see a small decrease of the error bars of $\Omega_m$ and $h$ by 9\% each.

We also show the combination of PantheonPLUS and CC data, where we observe a small increase of constraining power due to the PantheonPLUS, especially in the case without SF or RSD data. There is a mild synergy between the two added data sets, achieving a decrease in the error bars of around 9\% in $\Omega_m$ and 16\% in $h$ compared to the result when only including PantheonPLUS data. This is because the CC data prevent an upward shift in $\Omega_m$ and $h$ as seen in \cref{fig:pantheon}, caused by the slightly higher value of $\Omega_m$ in the PantheonPLUS data.

\enlargethispage*{2\baselineskip}
\subsection{The BAO+BBN probe with dark radiation}\label{ssec:neff}
Dark radiation can change the expansion rate in the early Universe, thus impacting the scale of the sound horizon (see \cref{eq:highredshift_expansion,eq:radiationdensity,eq:rs}). Due to the $H_0 - r_s$ degeneracy, the presence of dark radiation strongly affects the $H_0$ determination from both the CMB and the BAO+BBN probe. The latter, in particular, is sensitive to any additional  dark radiation at the time of the BBN, through the increased expansion rate, which in turns affect the light elements abundance. 

\begin{figure}[H]
    \centering
    \centerline{
    \includegraphics[width=0.54\textwidth]{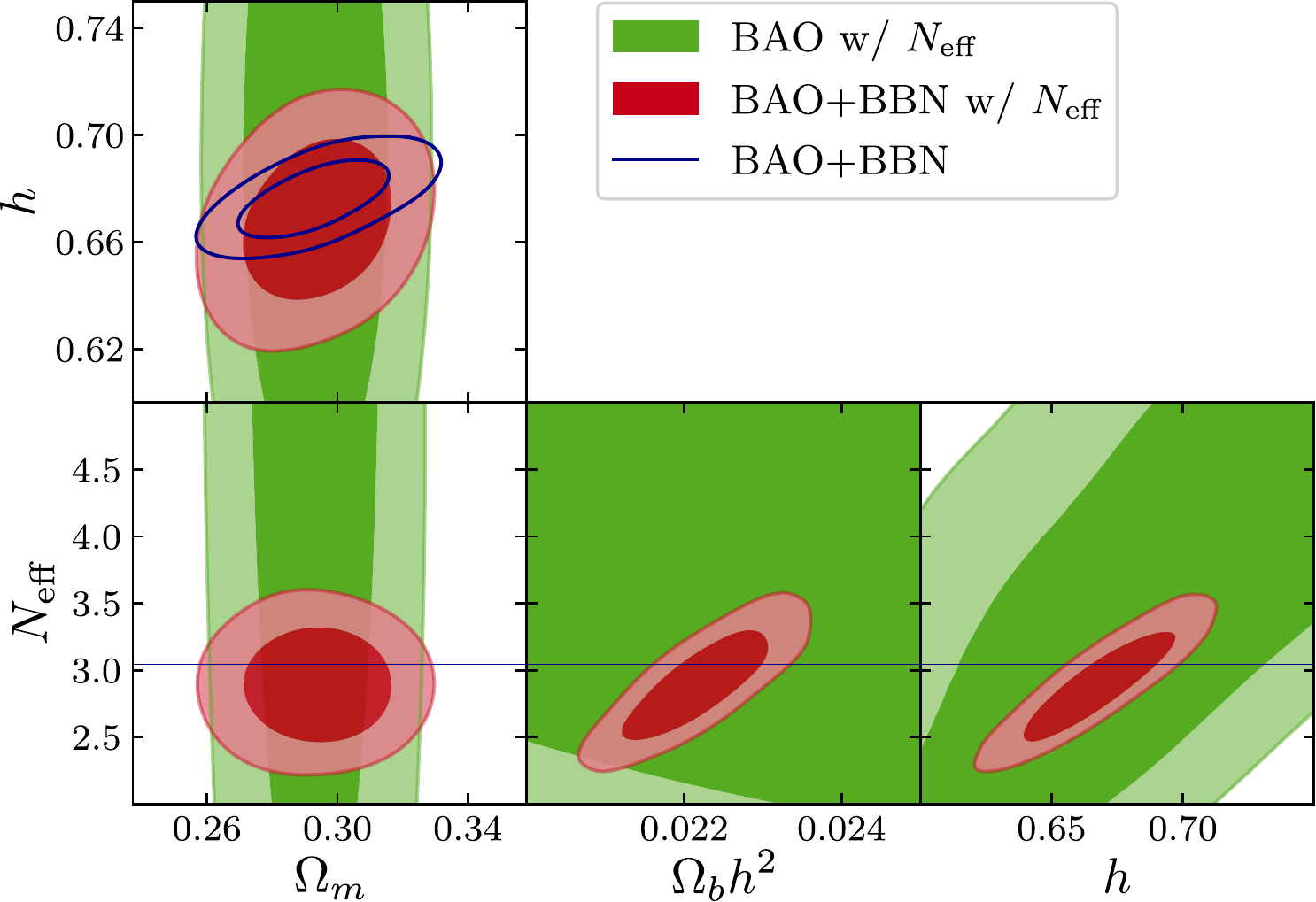}
    \includegraphics[width=0.54\textwidth]{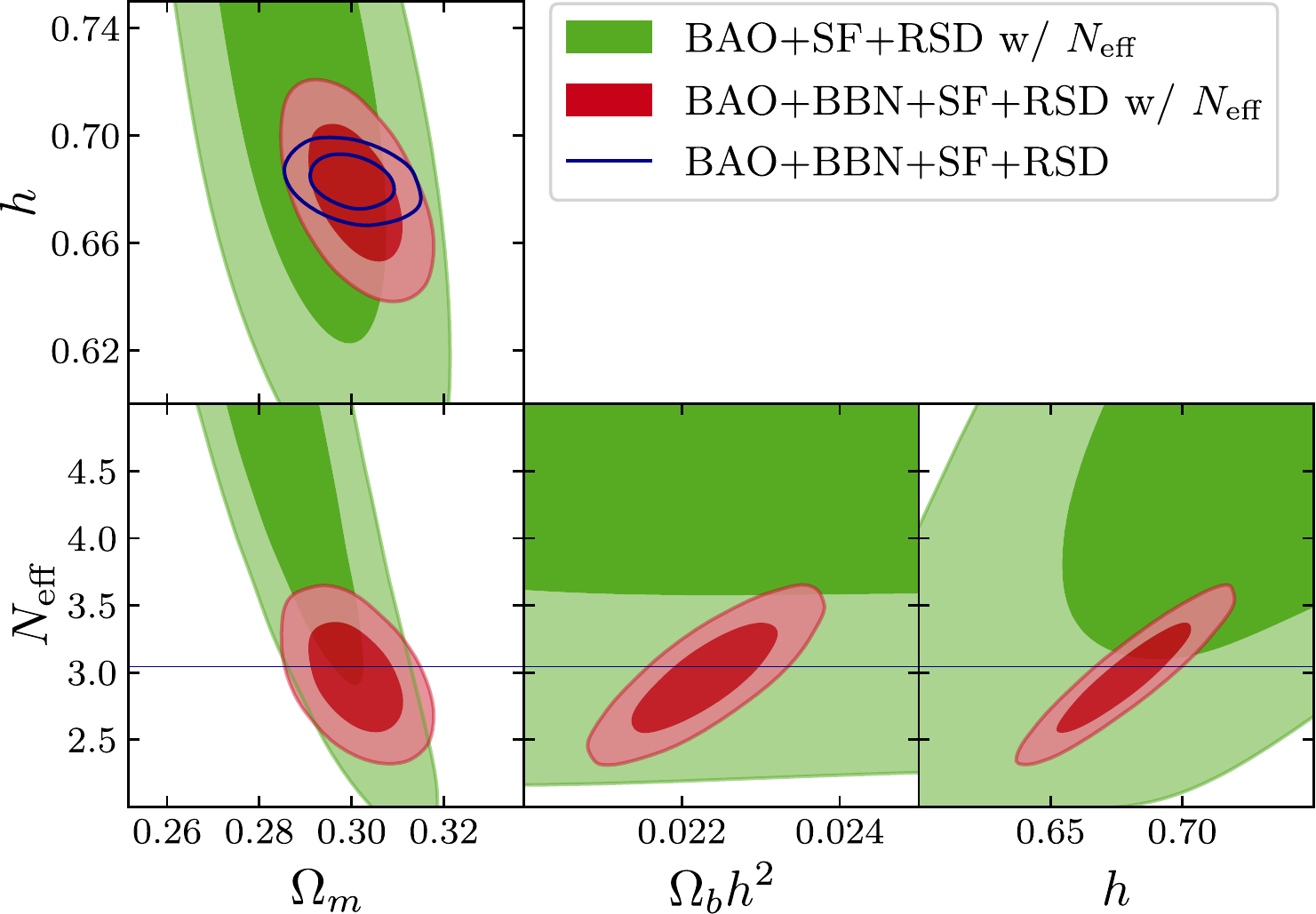}
    }
    \caption{Constraints of the model including additional dark radiation through $N_\mathrm{eff}$\,. SF denotes ShapeFit, and RSD denotes redshift space distortions. For reference, the solid blue lines correspond to the results in the $\Lambda$CDM model. \textbf{Left:} Without SF+RSD data. \textbf{Right:} With SF+RSD data.}
    \label{fig:neff}
\end{figure}

As such, the BAO+BBN probe can give independent and very relevant constraints for $N_\mathrm{eff}$ and $h$ in a way that does not depend on the precise Silk-damping mechanics of the dark radiation (which is very important for the CMB, see e.g. \cite{Schoneberg:2021qvd}). It is worth noting, however, that such constraints on $N_\mathrm{eff}$ can be avoided when considering a boost/production of the dark radiation abundance only after BBN. In \cref{fig:neff} we show the results both with and without BBN, in order to exemplify the importance of BBN in the constraints obtained for this model. 

For the BAO+BBN probe, the addition of $N_\mathrm{eff}$ does cause a widening of the constraints on the $\Omega_m-h$ plane, independently of whether SF/RSD data is included or not\footnote{Compared to the $\Lambda$CDM model, without SF+RSD the constraints widen by a factor of around $2$, to $H_0 = \cons{66.81}{1.95}{1.93} \mathrm{km/s/Mpc}$. Instead, with SF+RSD the constraints widen by a factor of around $2.5$, to $H_0 = \cons{67.89}{1.75}{1.80} \mathrm{km/s/Mpc}$.}. This is because the $H_0 - r_s$ degeneracy can only be partially be broken in this model\footnote{This is because there remains a $\{\Omega_b h^2, N_\mathrm{eff}\}$ degeneracy in BBN constraints, as an increase in $\Omega_b h^2$ can be partially compensated by an increase in $N_\mathrm{eff}$ while keeping the Deuterium abundance the same (see e.g. Fig 1 of \cite{Schoneberg:2019wmt}). Only when the $N_\mathrm{eff}$ becomes too large and starts to perturb the $Y_p$ measurement significantly, is this degeneracy broken, thus leading to overall much weaker constraints on $\Omega_b h^2$ in this model, allowing for additional freedom in $r_s$\,.}, allowing for only comparatively weaker constraints on $H_0$ after marginalizing over allowed values $N_\mathrm{eff}$ and the corresponding values of $r_s$\,. We note that the results are different from \cite{Brieden:2022lsd}, which used a BBN-motivated prior on $\Omega_b h^2$ instead of the full BBN likelihood used in this work, which primarily makes a difference for models of dark radiation that are constrained in this work by their impact on the $Y_p$ primordial abundance.

Instead, when abandoning the BBN constraints on this model altogether, both $\Omega_b h^2$ and $N_\mathrm{eff}$ become unconstrained, as expected from \cref{ssec:method}, leading to no constraint in $h$ (except due to prior bounds). Interestingly, the improved constraint of $\Omega_m$ from SF present in $\Lambda$CDM is also relaxed in the $N_\mathrm{eff}$ model without BBN. This is because the calibration of the $\Omega_m$ is of course an indirect result of the SF only possible when $\Omega_b h^2$ is sufficiently fixed (which it isn't without BBN), as otherwise the inherent degeneracies in the measurement of $m$ make precise inference of $h$ or $\Omega_m$ from SF alone impossible. Thus, only the \AP{}-based BAO measurement of $\Omega_m$ remains.

It is worth mentioning that also in this extended model RSD does not provide additional information compared to the BAO (the contours are identical to those displayed in the left panel of \cref{fig:neff}), as explained in \cref{ssec:shapefit}.

\begin{figure}[H]
    \centering
    \includegraphics[width=0.7\textwidth]{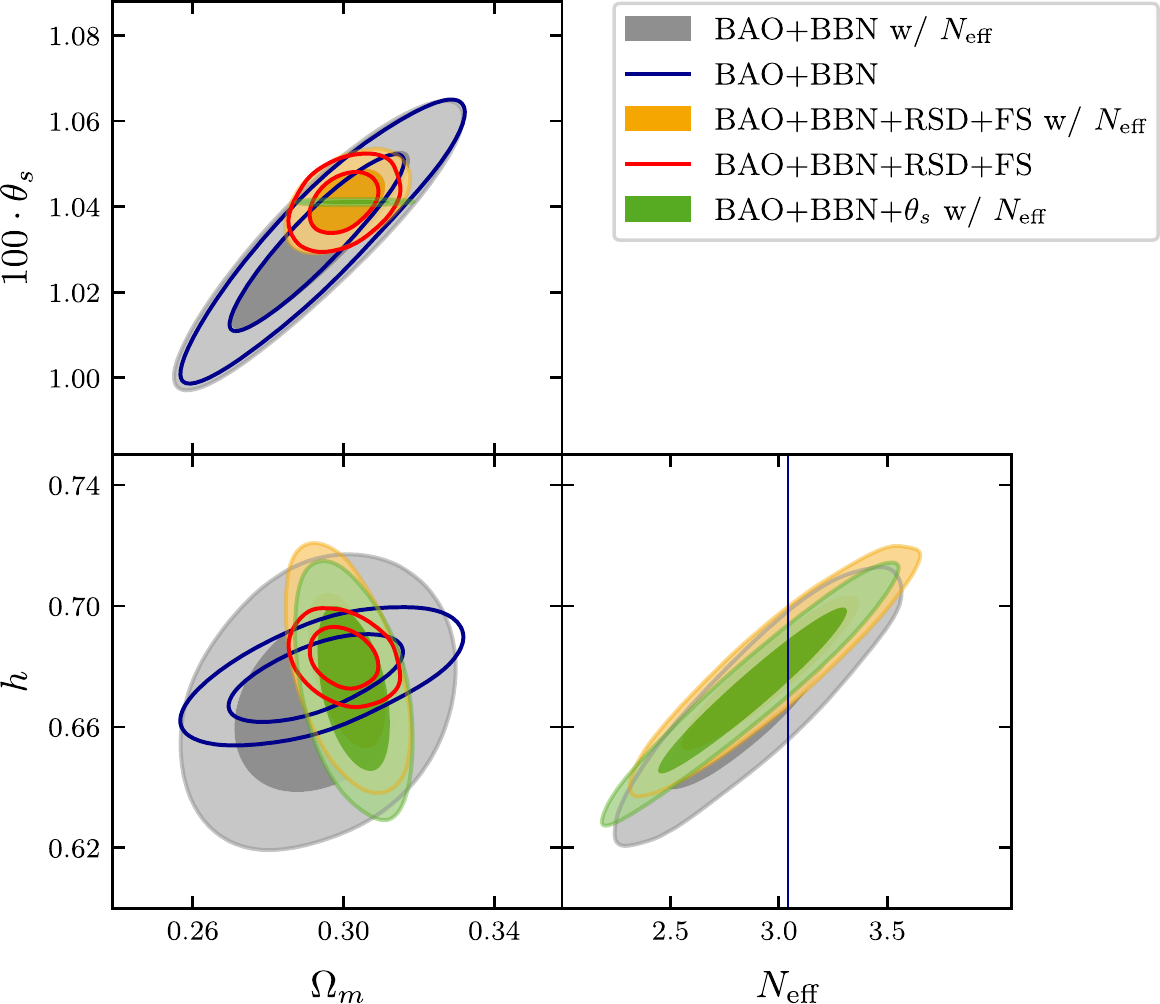}
    \caption{Constraints of the model including additional dark radiation through $N_\mathrm{eff}$ as well as a prior on $\theta_s$\, (green contour), compared to various other constraints (other contours). SF denotes ShapeFit, and RSD denotes redshift space distortions. For reference, the solid blue lines and the solid red lines correspond to the results in the $\Lambda$CDM model. The top left panel shows that the value of $100\cdot \theta_s$ inferred from the FS+RSD addition to BAO+BBN is very well in agreement with the sound horizon prior as measured from the CMB.}
    \label{fig:neff_thetas}
\end{figure}

In \cref{fig:neff_thetas} we also show that for the model including dark radiation one still obtains constraints of similar strength on the Hubble parameter $h$ when adding a angular sound horizon prior from the CMB compared to those of adding FS+RSD. More importantly, however, the BAO+BBN probe can -- in combination with the RSD+FS data -- constrain exactly this sound horizon angular scale very precisely (orange/red contour of \cref{fig:neff_thetas}) and the value inferred is very well in agreement with that from the CMB (green contour of \cref{fig:neff_thetas}), providing yet another crucial cross-check/consistency test. Applications of this cross-check to other models is left for future work.

\subsection{The BAO+BBN probe with shifted recombination}\label{ssec:me}
\begin{figure}[H]
    \centering
    \centerline{
    \includegraphics[width=0.50\textwidth]{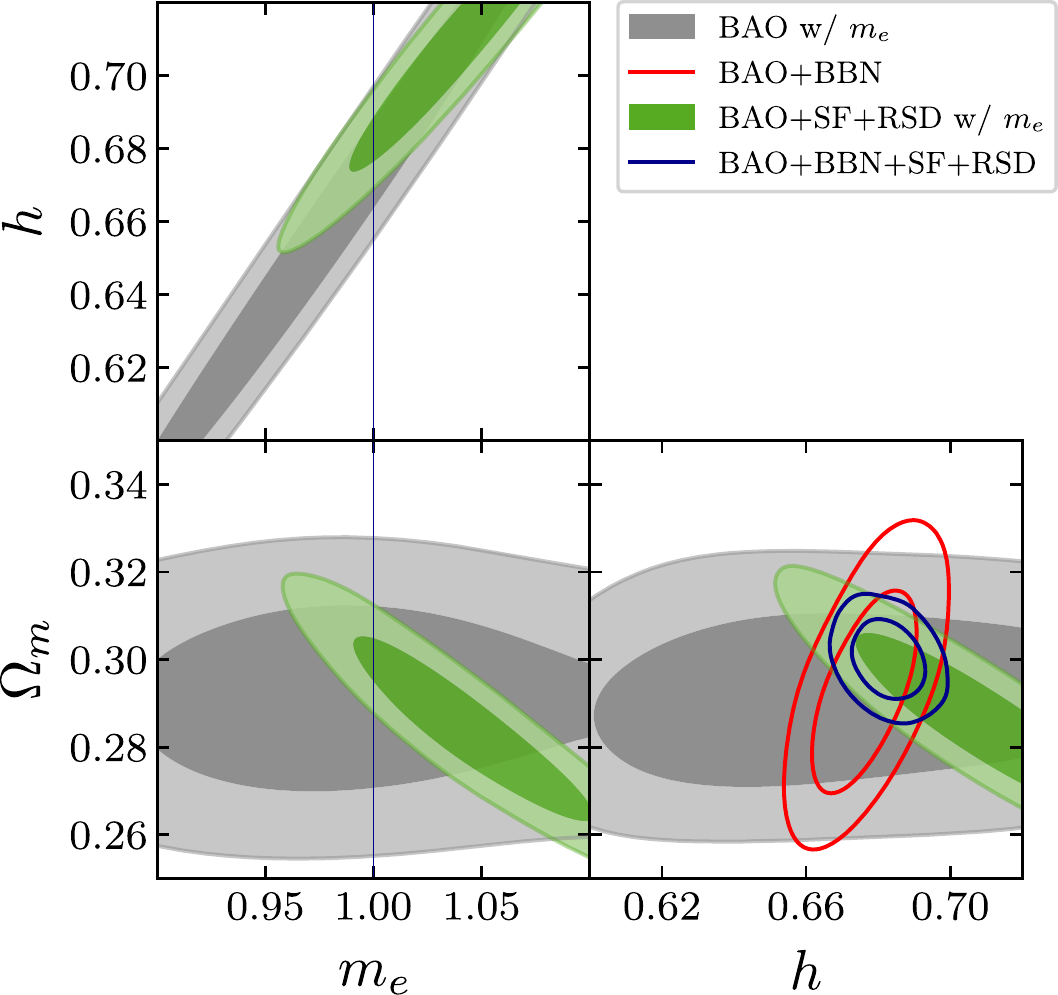}
    \includegraphics[width=0.55\textwidth]{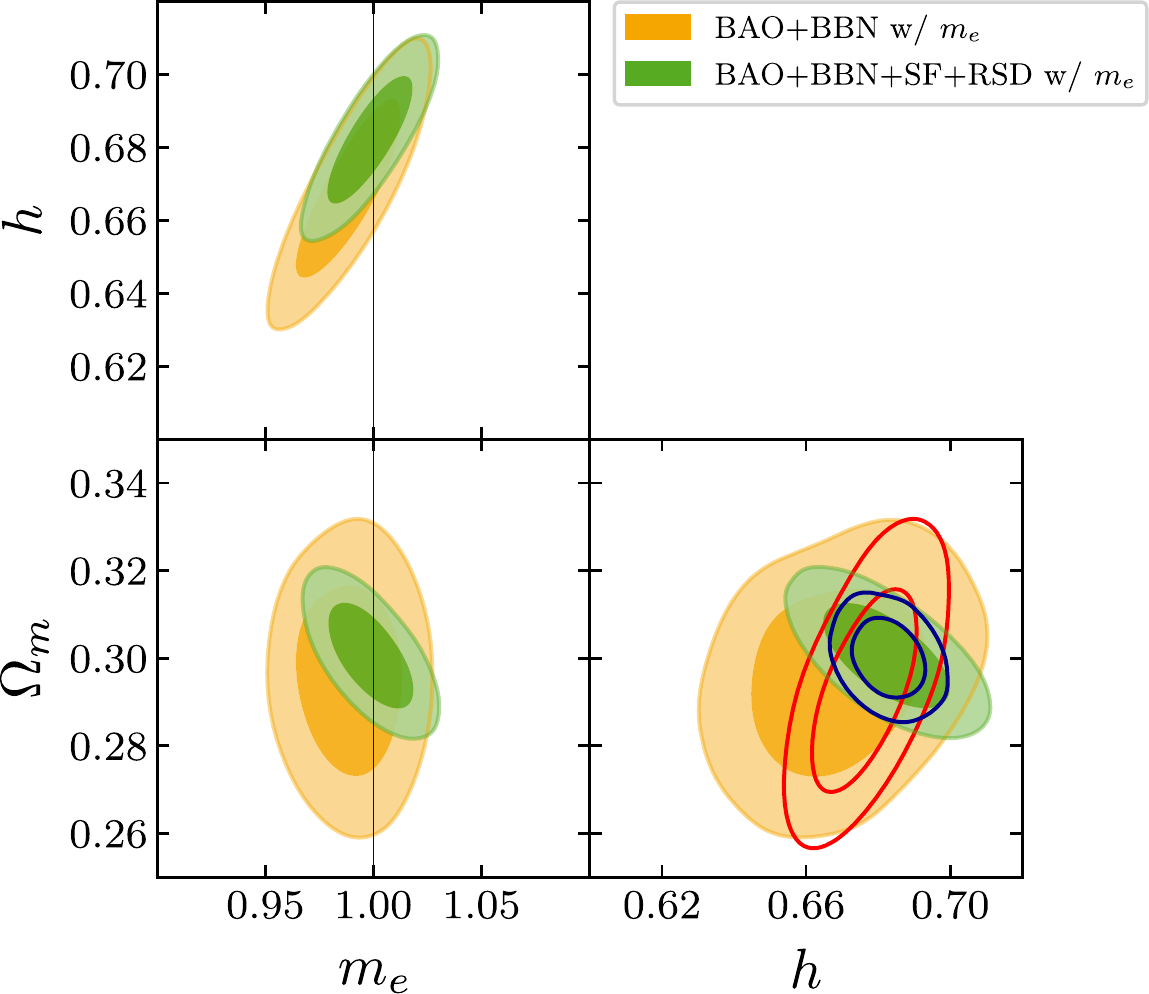}
    }
    \caption{Constraints for the BAO+BBN probe for the model of including additional variation of the redshift of recombination through a variation of the electron mass ($m_e$). The parameter $m_e$ is shown relative to the value measured today ($511$keV). SF denotes ShapeFit, and RSD denotes redshift space distortions. \textbf{Left:} Not including the impact of the model on BBN. \textbf{Right:} Including the impact of the model on BBN.}
    \label{fig:me}
\end{figure}

Another possibility to shift the sound horizon at recombination is by changing the redshift of recombination. While  several models  have been proposed,  in this work we focus on a model of a varying electron mass $m_e$ due to its simplicity. For this particular model, by investigating a few test cases, \cite{Seto:2022xgx} has shown that BBN can in principle provide tight constraints. Here we extend that discussion by performing a fully Bayesian analysis using the BBN likelihood of \cite{Schoneberg:2019wmt} with the prescription for including the electron mass as in \cite{Seto:2022xgx}.\footnote{This prescription only adjusts the neutron decay rate (see eq. 3 of \cite{Seto:2022xgx}), but this is expected to be the dominant effect on BBN. We leave a full implementation of the variation of other rates to future work.} The constraints are shown in the right panel of \cref{fig:me}.

However, we will also be more generally interested in treating this model as an effective toy model that produces shifts of the recombination redshift, which can also be accomplished by other mechanisms (such as \cite{Jedamzik:2020krr}). In this more general context, we will beging by  explicitly assuming the variation of the electron mass to be caused by a physical effect only after the times relevant for BBN. This will allow us to draw conclusions about models that shift the sound horizon of recombination without leaving an impact on BBN. We show the corresponding constraints in the left panel of \cref{fig:me}.

In the case that the variation does not leave an impact on BBN (left panel of \cref{fig:me}), we see that the constraints on the varying electron mass model are quite loose. We find that without SF or RSD data the Hubble parameter remains unconstrained, as the model-induced degeneracy between the varying electron mass and $h$ is perfect. Without the calibration of $r_s$ through BBN, the BAO+BBN probe cannot constrain $H_0$ anymore (except for the bounds from the prior). Instead, if SF data are added, weak constraints on $h$ are actually recovered. These stem from the $\{\omega_b,n_s\}$-calibrated measurement of the parameter $m$, which does allow for a joint measurement of approximately $\Omega_m h^{2} \approx \mathrm{const}$.
This demonstrates that, although currently not competitive with the BAO+BBN-derived constraints in $\Lambda$CDM, the SF does add valuable and independent information, which becomes especially important in extended models. We expect that such a result remains true for other extended models, but leave a more detailed investigation for future work.
This discussion is expected to also apply to models that shift the sound horizon through an increase in the expansion rate between BBN and recombination, which also shifts $r_s$ in a way that does not impact BBN, and thus leads to the same effects as discussed above. Note, however, that in this case we expect that  the interpretation of the SF parameter needs to be adjusted accordingly. We leave the detailed investigation of the SF constraint on such models modifying the expansion rate before recombination to future work.

In the case that we allow the varying $m_e$ - model to be constrained by BBN (right panel of \cref{fig:me}), we naturally find much tighter constraints. Indeed, we see that the constraints on $m_e$ are driven by BBN alone, and consequently   a similar indirect calibration of the sound horizon as in \cref{ssec:neff} is possible. In this case, the BAO+BBN probe by itself only gives relatively weak constraints on $h$. 

Similarly to \cref{ssec:neff} we find that the SF constraints in this case add a lot of additional constraining power due to their tight $\{\Omega_m,h\}$ correlation, though naturally the constraints are weaker than that of $\Lambda$CDM. Interestingly, they are still stronger than those in $\Lambda$CDM without SF (red line in \cref{fig:me}, right panel).

\section{Conclusions}\label{sec:conclusion}

The BAO+BBN probe has been shown to provide an excellent background probe of the expansion rate of the Universe. With the new results from the LUNA experiment \cite{Mossa:2020gjc} as well as the latest data release DR16 of SDSS \cite{eBOSS:2020yzd,SDSS-IV:2019txh} an update of the results is timely. More importantly, this gives us the occasion to investigate several additional aspects of this probe, such as combinations with a variety of additional measurements, and the robustness to the choice of the underlying cosmological model.

First and foremost, we find that the new data put very tight constraints on the Hubble parameter without relying on CMB data: $H_0 = \cons{67.6}{0.9}{1.0} \mathrm{km/s/Mpc}$ in $\Lambda$CDM, which breaches the 1km/s/Mpc mark recently achieved by local distance ladder determinations \cite{Riess:2021jrx}. This represents a tightening of the constraint by roughly 20\% and a $0.6\sigma$ downward shift from the result in \cite{Schoneberg:2019wmt}, $H_0 = \cons{68.3}{1.1}{1.2} \mathrm{km/s/Mpc}$. The Gaussian tension (GT) with the newest PantheonPLUS+SH0ES result \cite{Riess:2021jrx} has thus grown from $3.2\sigma$ to $3.7\sigma$.

\pagebreak[40]
Secondly, there is no longer any reliance on datasets that carry a mild internal tension at the $2 \sigma$ level. It is worth noting that the increased precision is possible even without the highest redshift data used in this work (Ly$\alpha$ from DR14) due to the statistical power of the DR16 measurement of the BAO in QSO already providing sufficient lever arm to break the degeneracy between $H_0 r_s$ and $\Omega_m$\,. We explicitly mention that we do not find any discrepancy of the Ly$\alpha$ data from the other data sets.

Third, we investigated the addition of other features that can be extracted from the clustering data, such as the redshift space distortion (RSD) constraints on the parameter combination $f \sigma_8$ or the ShapeFit (SF) constraints on the parameter $m$. Especially the latter is very synergistic with the BAO+BBN measurement (if a prior on $n_s$ is included), leading to competitively tight constraints at the level of $H_0 = 68.3 \pm 0.7 \mathrm{km/s/Mpc}$ ($3.8 \sigma$ GT). Instead, RSD does not add a lot of synergistic information to the BAO+BBN measurement in terms of the Hubble tension, and leads to essentially the same constraints as without (even with a prior on $A_s$ imposed).

Fourth, we investigate how the constraints from the BAO+BBN probe improve if other synergistic geometrical background data are added to it. The early Universe angular sound horizon of the BAO observed in the CMB ($\theta_s$) is measured at sub-percent precision (0.03\%) in an almost model-independent way. With only this additional information from the CMB, the precision of the BAO+BBN probe reaches the same level as the original CMB+CMB lensing+BAO analysis in \cite{Planck:2018vyg}, reaching $H_0 = 68.2\pm 0.45\mathrm{km/s/Mpc}$ ($4.2\sigma$ GT). More importantly, the region in parameter space preferred by this combination is perfectly in agreement with that preferred by adding the SF measurement. If we instead add Pantheon data, we also notice a slight increment of the constraining power, comparable to that when the SF data are added. Since the newest PantheonPLUS \cite{Brout:2022vxf} has a slightly different bestfit ($\sim 1.9 \sigma$) for $\Omega_m$ from the BAO data, the results are different compared to the addition of the older Pantheon data \cite{Pan-STARRS1:2017jku}, leading either to $H_0 = \cons{67.8}{0.7}{0.8} \mathrm{km/s/Mpc}$ (Pantheon, $4.2\sigma$ GT) or $H_0 = \cons{68.3}{0.9}{1.0}$ (PantheonPLUS, $3.5\sigma$ GT). The addition of Cosmic Chronometer (CC) data decrease the uncertainties slightly (by $\sim 9\%$), and does prefer the same region in parameter space, again demonstrating the nice consistency between the probes.

Finally, we investigate the model-dependence of the BAO+BBN constraints with respect to early-time solutions of the Hubble tension. We show that models of dark radiation or a varied electron mass are constrained by BBN. The resulting constraints are degraded only by a factor of 2 compared to the $\Lambda$CDM model (and are tighter than when only a $\Omega_b h^2$ prior is included as in \cite{Brieden:2022lsd}). This is expected to generalize for other models that can be constrained with BBN.
A comparison with \cite{Smith:2022iax} highlights this strong impact of the BBN probe on the allowed parameter regions of the $N_\mathrm{eff}$ and $m_e$ models: Upon inclusion of BBN constraints we find that the Hubble tension with the SH0ES results increases to $2.8\sigma$ GT ($N_\mathrm{eff}$) and $3.5\sigma$ GT ($m_e$), respectively.
On the other hand, models that purely shift the sound horizon at recombination without measurable impact on the BBN are expected not to be bounded by the BAO+BBN probe, though the additional information from SF provides hope for constraints in the near future.

We conclude that the BAO+BBN combination is an incredibly interesting and constraining probe of the geometrical expansion history of the universe and the Hubble tension, especially when combined with the ShapeFit measurement as described in \cite{Brieden:2021edu,Brieden:2022lsd} to give a $0.7\mathrm{km/s/Mpc}$ determination of the Hubble parameter. Even further, the BAO+BBN probe profits quite dramatically from combinations with other geometric probes, such as the angular sound horizon of the CMB, uncalibrated standardized supernovae of type Ia, or Cosmic Chronometers. In particular, all inferences of the Hubble parameter from these various combinations agree on the inferred Hubble parameter and agree with the determination from the CMB anisotropies. Finally, we show that with a word of caution there is a dependence of the BAO+BBN probe on the assumed underling model, if it modifies the early-time expansion history. While constraints can be obtained for models that have an impact on the BBN, no inference on the Hubble parameter can be made for other models. However, we find a hint that the addition of a precise shape measurement could provide important independent constraints on these models.

In the future new data from ongoing and upcoming large scale structure surveys will most definitely make the BAO+BBN probe together with its SF and RSD additions one of the most interesting cosmological probes of modern cosmology, providing a crucial cross-check of the results derived from the CMB and the local distance ladder. This will also allow to shed new light on the proposed theoretical solutions to the Hubble tension.

\acknowledgments

We would like to thank Vivian Poulin for providing the PantheonPLUS likelihood and helpful discussions. We would further like to thank Raul Jimenez for pointing out several works providing a critical discussion on Lyman-$\alpha$ based BAO data, sparking the idea for \cref{ssec:internaltension} as a response. We would also like to thank Julien Lesgourgues for helpful comments and a detailed discussion on BBN theoretical uncertainties. Finally, we would like to thank Guillermo Franco Abell\'an for his comments. Funding for this work was  provided by project PGC2018-098866-B-I00
MCIN/AEI/10.13039/501100011033 y FEDER “Una manera de hacer Europa”, as well as the
“Center of Excellence Maria de Maeztu 2020-2023” award to the ICC University Barcelona (CEX2019-000918-M funded by MCIN/AEI/10.13039/501100011033) and European Union’s Horizon 2020 research and innovation programme ERC (BePreSysE, grant agreement 725327).


\bibliography{bib}
\bibliographystyle{JHEP}

\appendix
\newpage
\section{Supplemental data}
In \cref{tab:results} we summarize all the resulting constraints in $H_0$ and $\Omega_m$ as shown in the corresponding triangle plots in section \cref{sec:results}.
\begin{table}[h]
    \centering
    \begin{tabular}{l|r r}
Run name & $H_0 [\mathrm{km/s/Mpc}]$ & $\Omega_m$\\ \hline \hline
$\Lambda$CDM model&&\\ \hline
BAO+BBN & $67.64^{+0.94}_{-1.03}$ & $0.293^{+0.015}_{-0.016}$\\
BAO+BBN+SF+RSD & $68.29^{+0.68}_{-0.69}$ & $0.3002^{+0.0060}_{-0.0063}$\\ \hline
BAO(DR12)+BBN(noLUNA) & $68.36^{+1.13}_{-1.25}$ & $0.302^{+0.018}_{-0.020}$\\
BAO+BBN(noLUNA) & $67.90^{+0.92}_{-1.03}$ & $0.294^{+0.015}_{-0.016}$\\
BAO(DR12)+BBN & $68.14^{+1.13}_{-1.24}$ & $0.302^{+0.017}_{-0.020}$\\ \hline
BAO(QSO)+BBN & $64.51^{+2.40}_{-2.69}$ & $0.347^{+0.049}_{-0.087}$\\
BAO(LRG)+BBN & $71.18^{+2.27}_{-2.41}$ & $0.349^{+0.037}_{-0.035}$\\
BAO(QSO+LRG)+BBN & $67.88^{+1.10}_{-1.46}$ & $0.299^{+0.018}_{-0.025}$\\
BAO(QSO+Ly$\alpha$)+BBN & $66.15^{+2.22}_{-2.24}$ & $0.302^{+0.035}_{-0.041}$\\
BAO(DR12 QSO+LRG)+BBN & $72.62^{+2.75}_{-3.56}$ & $0.375^{+0.043}_{-0.052}$\\ \hline
BAO+BBN+RSD & $67.66^{+0.90}_{-0.98}$ & $0.29^{+0.015}_{-0.016}$\\
BAO+BBN+SF & $68.30^{+0.66}_{-0.69}$ & $0.3023^{+0.0061}_{-0.0063}$\\
BAO+BBN+SF+RSD (no $n_s$ prior) & $67.70^{+0.91}_{-0.93}$ & $0.288^{+0.014}_{-0.013}$\\
BAO+BBN+RSD (+$A_s$ prior) & $67.88\pm0.96$ & $0.293^{+0.017}_{-0.013}$\\
BAO+BBN+SF+RSD (+$A_s$+$n_s$ prior) & $68.25^{+0.88}_{-0.89}$ & $0.299^{+0.013}_{-0.014}$\\
BAO+BBN+SF+RSD (+$A_s$ prior) & $68.47\pm0.68$ & $0.3001\pm0.0062$\\ \hline
BAO+BBN+$\theta_s$ & $68.16^{+0.48}_{-0.49}$ & $0.3022^{+0.0062}_{-0.0064}$\\
BAO+BBN+SF+RSD+$\theta_s$ & $68.30\pm0.45$ & $0.3001^{+0.0055}_{-0.0060}$\\ \hline
BAO+BBN+Pantheon & $67.75^{+0.70}_{-0.84}$ & $0.296^{+0.011}_{-0.010}$\\
BAO+BBN+PantheonPLUS & $68.29^{+0.86}_{-0.97}$ & $0.311^{+0.011}_{-0.012}$\\
BAO+BBN+SF+RSD+Pantheon & $68.34^{+0.66}_{-0.65}$ & $0.3001^{+0.0057}_{-0.0062}$\\
BAO+BBN+SF+RSD+PantheonPLUS & $68.20^{+0.67}_{-0.69}$ & $0.3036^{+0.0055}_{-0.0062}$\\ \hline
BAO+BBN+CC & $68.12^{+0.90}_{-0.88}$ & $0.302\pm0.014$\\
BAO+BBN+SF+RSD+CC & $68.19^{+0.69}_{-0.68}$ & $0.3002^{+0.0061}_{-0.0064}$\\
BAO+BBN+PantheonPLUS+CC & $68.28^{+0.73}_{-0.81}$ & $0.311^{+0.01}_{-0.011}$\\
BAO+BBN+SF+RSD+PantheonPLUS+CC & $68.20^{+0.67}_{-0.68}$ & $0.3042^{+0.0058}_{-0.0060}$\\ \hline \hline
$N_{\rm eff}\Lambda$CDM model & & \\\hline

BAO w/ $N_\mathrm{eff}$ & --- & $0.293\pm0.015$\\
BAO+BBN w/ $N_\mathrm{eff}$ & $66.81^{+1.95}_{-1.93}$ & $0.294^{+0.016}_{-0.013}$\\
BAO+SF w/ $N_\mathrm{eff}$ & $73.12^{+8.31}_{-6.48}$ & $0.288^{+0.009}_{-0.013}$\\
BAO+BBN+SF w/ $N_\mathrm{eff}$ & $67.89^{+1.68}_{-1.81}$ & $0.3009\pm0.0070$\\ \hline \hline
$m_e\Lambda$CDM model & & \\ \hline
BAO w/ $m_e$ & --- & $0.291\pm0.014$\\
BAO+SF+RSD w/ $m_e$ & $70.82^{+2.81}_{-2.09}$ & $0.286^{+0.011}_{-0.016}$\\
BBN w/ $m_e$ & --- &  ---\\
BAO+BBN w/ $m_e$ & $66.67^{+1.47}_{-1.83}$ & $0.298^{+0.014}_{-0.011}$\\
BAO+BBN+SF+RSD w/ $m_e$ & $68.21^{+1.12}_{-1.18}$ & $0.3009^{+0.0079}_{-0.0080}$\\
    \end{tabular}
    \caption{Mean and $1\sigma$ uncertainty for $H_0$ and $\Omega_m$ for all runs, labeled according to the shorthand notation introduced in \cref{ssec:data} and used in \cref{sec:results}.\vspace*{-2\baselineskip}}
    \label{tab:results}
\end{table}
\end{document}